\documentclass[useAMS,usenatbib,fleqn]{mn2e}
\usepackage{times} 
\usepackage{listings}
\usepackage{graphics}
\usepackage{subfigure}
\usepackage{graphicx}
\usepackage{graphicx,xspace}
\usepackage{amsmath}
\usepackage{amssymb}
\usepackage{hyperref}
\usepackage{aas_macros}
\usepackage{lscape}
\usepackage{rotating}
\usepackage{placeins}
\usepackage{natbib}
\usepackage{color}

\graphicspath{{./Paper_plots/}}
\pdfpageattr {/Group << /S /Transparency /I true /CS /DeviceRGB>>}

\newcommand{\galform}{{\sc{galform}}\xspace}
\newcommand{\grasil}{{\sc{grasil}}\xspace}
\newcommand{\magphys}{{\sc{magphys}}\xspace}

\def\apjl{ApJL}

\def\aap{A\&A}

\def\apjs{ApJS}

\title[The clustering evolution of dusty star-forming galaxies]{The clustering evolution of dusty star-forming galaxies}
\author[W. I. Cowley et al.]{
\parbox[t]{\textwidth}{
William I. Cowley\thanks{E-mail: w.i.cowley@durham.ac.uk }, 
Cedric G. Lacey,
Carlton M. Baugh,
Shaun Cole}
\vspace*{6pt}\\
Institute for Computational Cosmology, Department of Physics,
University of Durham, South Road, Durham, DH1 3LE, UK.
\vspace*{-0.5cm}}

\begin{document}
\date{\today}
\pagerange{\pageref{firstpage}--\pageref{lastpage}} \pubyear{2015}
\maketitle
\label{firstpage}
\begin{abstract}
We present predictions for the clustering of galaxies selected by their emission at far infra-red (FIR) and sub-millimetre wavelengths.  This includes the first predictions for the effect of clustering biases induced by the coarse angular resolution of single-dish telescopes at these wavelengths.  We combine a new version of the \galform model of galaxy formation with a self-consistent model for calculating the absorption and re-emission of radiation by interstellar dust.  Model galaxies selected at $850$~$\mu$m reside in dark matter halos of mass $M_{\rm halo}\sim10^{11.5}-10^{12}$~$h^{-1}$~M$_{\odot}$, independent of redshift (for $0.2\lesssim z\lesssim4$) or flux (for $0.25\lesssim S_{850\mu\rm m}\lesssim4$~mJy).  At $z\sim2.5$, the brightest galaxies ($S_{850\mu\rm m}>4$~mJy) exhibit a correlation length of $r_{0}=5.5_{-0.5}^{+0.3}$~$h^{-1}$~Mpc, consistent with observations.  We show that these galaxies have descendants with stellar masses $M_{\star}\sim10^{11}$~$h^{-1}$~M$_{\odot}$ occupying halos spanning a broad range in mass $M_{\rm halo}\sim10^{12}-10^{14}$~$h^{-1}$~M$_{\odot}$.  The FIR emissivity at shorter wavelengths ($250$, $350$ and $500$~$\mu$m) is also dominated by galaxies in the halo mass range $M_{\rm halo}\sim10^{11.5}-10^{12}$~$h^{-1}$~M$_{\odot}$, again independent of redshift (for $0.5\lesssim z\lesssim5$).  We compare our predictions for the angular power spectrum of cosmic infra-red background anisotropies at these wavelengths with observations, finding agreement to within a factor of $\sim2$ over all scales and wavelengths, an improvement over earlier versions of the model. Simulating images at $850$~$\mu$m, we show that confusion effects boost the measured angular correlation function on all scales by a factor of $\sim4$.  This has important consequences, potentially leading to inferred halo masses being overestimated by an order of magnitude.
\end{abstract}


\section{Introduction}
\label{sec:Introduction}

The emission from galaxies formed throughout cosmic history appears as a diffuse cosmological background.  The far infra-red (FIR) and sub-millimetre (sub-mm, $100$~$\mu$m - $1$~mm) part of this background \citep[CIB, e.g.][]{Puget96,Fixsen98} is mostly produced by the re-emission of stellar radiation by interstellar dust, with a small ($\lesssim5\%$) contribution from dust heated by UV/X-ray emission from AGN \citep[e.g.][]{Almaini99,Silva04}, and has a similar energy density to the background at UV/optical wavelengths \citep[e.g.][]{HauserDwek01,Dole06}.  This implies that most of the star formation over the history of the Universe has been obscured by dust.  Understanding the nature of the galaxies that contribute to the CIB is therefore critical to a full understanding of galaxy formation.  

Much progress has been made in recent years to map the sky at these long wavelengths either from space, using satellites such as the \emph{Herschel} Space Observatory \citep{Pilbratt10}, its predecessor the Balloon-borne Large Aperture Sub-millimetre Telescope \citep[BLAST; ][]{Devlin09}, and the \emph{Planck} Satellite\footnote{\url{http://www.esa.int/Planck}}, or from ground-based instruments, such as the Super Common User Bolometer Array 2 \citep[SCUBA-2; ][]{Holland13}.  However, one of the key problems with observations at these long wavelengths is confusion noise, caused by the coarse angular resolution [$\sim20$~arcsec full width at half maximum (FWHM)] of the telescopes and the high surface density of detectable objects.  This means that only the brightest objects can be resolved above the confusion background from imaging at these wavelengths.  

Whilst these individually resolved galaxies do not form the dominant contribution to the CIB \citep[e.g.][]{Oliver10}, they are important to study in their own right as they appear to be amongst the most highly star-forming objects in the Universe, as their FIR/sub-mm emission is thought to be powered by star formation, leading to inferred star formation rates (SFRs) of $\gtrsim 100$~M$_{\odot}$~yr$^{-1}$ \citep[e.g.][]{Smail02,Michalowski10,Swinbank13}.  However, the use of gravitational lensing \citep[e.g.][]{Smail97,Knudsen08,Chen13}, stacking techniques (e.g. B\'{e}thermin et al. \citeyear{Bethermin12}; Geach et al. \citeyear{Geach13}) and interferometers \citep[e.g.][]{Hatsukade13,Carniani15} has to some extent circumvented the problem of confusion noise and allowed up to $\sim80\%$ of the CIB to be statistically resolved into galaxies.   

Placing these FIR/sub-mm galaxies into a consistent evolutionary context has proven challenging.  In terms of resolved sub-mm galaxies (SMGs) it is still unclear what physical mechanism triggers the prodigious SFRs inferred from observations. In the local Universe ($z\lesssim0.3$), the majority of ultra-luminous galaxies ($L_{\rm IR}>10^{12}$~L$_{\odot}$) are gas-rich major mergers (e.g. Sanders \& Mirabel \citeyear{SandersMirabel96}), but whether this is the dominant triggering mechanism at the peak of the SMG redshift distribution \citep[$z\sim2.5$, e.g.][]{Chapman05,Simpson13} is unclear.  Some dynamical studies using emission lines from the \textsuperscript{12}CO molecule \citep[e.g.][]{Tacconi08} and H$\alpha$ \citep[e.g.][]{MenendezDelmestre13} have concluded that they see evidence of merger activity, though the sample sizes are small ($\lesssim10$ objects).  The merger scenario is also supported by some recent morphological studies \citep[e.g.][]{Chen15}.  However, examples of rotationally supported discs have also been found \citep[e.g.][]{Swinbank11} suggesting the star formation was triggered by secular disc instabilities.  Simulations suggest that the contraction of gas towards the centre of a galaxy, fuelling the star formation which results in the enhanced FIR/sub-mm emission \citep[e.g.][]{MihosHernquist96,Chakrabarti08,Narayanan10}, could also cause accretion onto a supermassive black hole, with the resulting quasar phase quenching the star formation \citep[e.g.][]{DiMatteo05}, possibly resulting in compact quiescent galaxies \citep[e.g.][]{Toft14}.  It has been speculated that SMGs could then evolve onto the scaling relations observed for massive local elliptical galaxies, based on simple arguments involving the timescale of the burst and the ageing of the stellar population \citep[e.g.][]{Lilly99,Swinbank06,Simpson13}, and assuming that most of the stellar mass at $z=0$ is put in place during the `SMG phase'.  However, \cite{Gonzalez11} present an alternative scenario in which SMGs evolve into galaxies with stellar mass $\sim10^{11}$ $h^{-1}$~M$_{\odot}$ at $z=0$, with the SMG phase accounting for little of this stellar mass.    

An important constraint on any evolutionary picture can come from observational measurements of the clustering of selected galaxies, which provides information on the masses of the dark matter halos in which they reside. However, measuring the clustering of FIR/sub-mm galaxies has proven challenging. Some studies have failed to produce significant detections of clustering \citep[e.g.][]{Scott02,Webb03,Coppin06,Williams11}, or the results derived from similar data have proven contradictory \citep[e.g.][]{Cooray10,Maddox10}.  Nevertheless, at $850$~$\mu$m \cite{Hickox12} used a cross-correlation analysis \citep[e.g.][]{Blake06} to find that SMGs selected from the LESS\footnote{Large APEX (Atacama Pathfinder EXperiment) Bolometer Camera Array (LABOCA) Extended \emph{Chandra} Deep Field South (ECDFS) Sub-millimetre Survey} source catalogue \citep{Weiss09} have a correlation length of $r_{0}=7.7_{-2.3}^{+1.8}$ $h^{-1}$~Mpc.  This result is consistent with an earlier study by \cite{Blain04} who used a pair-counting analysis to show that SMGs selected from a number of SCUBA fields have a correlation length of $6.9\pm2.1$ $h^{-1}$~Mpc.  These correlation lengths are consistent with SMGs residing in halos of mass $10^{12}-10^{13}$~$h^{-1}$~M$_{\odot}$.  Both the Hickox et al. and Blain et al. studies were performed prior to interferometric observations, which showed that many single-dish sources are in fact composed of multiple, fainter galaxies \citep[e.g.][]{Wang11,Hodge13}.  It is currently unclear from previous work how this result affects the observed clustering of SMGs.  We therefore present predictions for this in Section~\ref{sec:angular}.

Information about the clustering, and therefore host halo masses, of the unresolved FIR/sub-mm galaxies which contribute to the bulk of the CIB, can be obtained from the angular power spectrum of CIB anisotropies.  The first attempts to measure this, by \cite{Peacock00} for the Hubble Deep Field observed by SCUBA at $850$~$\mu$m, and \cite{Lagache00} for a $0.25$~deg$^2$ \emph{Infrared Space Observatory} (ISO) field at $170$~$\mu$m, found at best only a tentative signal above the shot noise. More recently studies have been able to measure a clear signal (e.g. Viero et al. \citeyear{Viero09}; Amblard et al. \citeyear{Amblard11}; Viero et al. \citeyear{Viero13}; Planck Collaboration XXX \citeyear{Planck14CIB}), though significant modelling is required in order to interpret these results in terms of halo masses. The Viero et al. (2013) and Plank Collaboration studies infer the typical halo mass for galaxies that dominate the CIB power spectrum as $10^{11.95\pm0.5}$~$h^{-1}$~M$_{\odot}$ and $10^{12.43\pm0.1}$~$h^{-1}$~M$_{\odot}$ respectively, making various assumptions such as the form of the relationship between galaxy luminosity and halo mass being independent of redshift,  and that this relationship is the same for both central and satellite galaxies.

Historically, hierarchical models of galaxy formation have struggled to simultaneously match the number density of FIR/sub-mm galaxies at high redshift $(z\gtrsim2)$ and the present day ($z=0$) luminosity function in optical and near-IR bands \cite[e.g.][]{Granato00}.  It follows that theoretical predictions for the clustering, and host halo masses, of such galaxies are few. \cite{VanKampen05} present a number of predictions for the angular clustering of SMGs under different scenarios. However, these models are phenomenological and do not attempt to predict the sub-mm flux of galaxies in a self-consistent manner.  \cite{Baugh05} presented a version of \galform, the Durham semi-analytic model of hierarchical galaxy formation \citep{Cole00}, which successfully reproduced the observed number counts and redshift distribution of SMGs at $850$~$\mu$m as well as the $z=0$ luminosity function in optical and near infra-red bands.  In order to do so these authors found it necessary to dramatically increase the importance of high redshift galaxy mergers relative to earlier versions of \galform \citep[e.g.][]{Cole00,Benson03} through the introduction of a top-heavy IMF in starburst galaxies.  In this instance sub-mm flux was calculated by combining \galform with the radiative transfer code \grasil \citep{Silva98}.  Predictions of the SMG clustering in this model were presented in \cite{Almeida11}, who found a correlation length of $5.6\pm0.9$ $h^{-1}$~Mpc for galaxies with $S_{850 \mu\rm m}>5$~mJy at $z=2$, in good agreement with the subsequent observational measurement of \cite{Hickox12}.  The angular power spectrum of CIB anisotropies predicted by this model was presented in \cite{Hank12} and was within a factor of $\sim3$ of the measurements of the Planck Collaboration (XVIII, \citeyear{Planck11CIB}).

Predictions for the clustering of FIR/sub-mm selected galaxies from hydrodynamical simulations of galaxy formation are limited due to the relatively small volumes that can (currently) be simulated using this method $\sim (100$ $h^{-1}$~Mpc)$^3$ \citep[e.g.][]{Vogelsberger14,Schaye15} and the computational expense of the radiative transfer required to properly calculate the sub-mm fluxes of the simulated galaxies.  Nevertheless, \cite{Dave10} used a hydrodynamical simulation to argue that $850$~$\mu$m SMGs at $z=2$ should be a highly clustered population with a correlation length of $r_{0}\sim10$ $h^{-1}$~Mpc and a bias of $\sim6$.  However, this work did not calculate the sub-mm flux for any of the simulated galaxies and instead relied entirely on the ansatz that SMGs are the most highly star-forming galaxies at a given epoch, with a SFR selection limit chosen such that the number density of the simulated sample matched that of observed SMGs. 

Here we present predictions for the clustering, and host halo masses, of galaxies selected by total infra-red luminosity, and FIR/sub-mm emission.  We use a new version of the \galform semi-analytic model \citep[][henceforth L15]{Lacey15}.  This is combined with a simple model for the reprocessing of stellar radiation by dust in which the dust temperature is calculated self-consistently \citep[as is done in e.g.][]{Gonzalez11,Hank12}.  This paper is structured as follows: in Section~\ref{sec:Model} we introduce the theoretical model, in Section~\ref{sec:spatial} we present predictions for the spatial clustering of galaxies selected by their total infra-red luminosity ($L_{\rm IR}$), and by their $850$~$\mu$m flux, in Section~\ref{sec:angular} we make predictions for the angular clustering of simulated galaxies selected by their $850$~$\mu$m flux, taking into account the effect of the single-dish beam used to make such observations, and in Section~\ref{sec:Pk} we present predictions for the angular power spectrum of CIB anisotropies at $250$, $350$, and $500$~$\mu$m.  We conclude in Section~\ref{sec:summary}.         

\section{The Theoretical Model}
Here we introduce our model, which combines a dark matter only $N$-body simulation, a state-of-the-art semi-analytic model of galaxy formation and a simple model for the reprocessing of stellar radiation by dust in which the dust temperature is calculated self-consistently based on radiative transfer and global energy balance arguments.  We also briefly describe some of the physical properties of the dusty star-forming galaxies in this model.

\label{sec:Model}
\subsection{GALFORM}

The Durham semi-analytic model of hierarchical galaxy formation, \galform, was introduced in \cite{Cole00}, building on ideas outlined by \cite{WhiteRees78}, \cite{WhiteFrenk91} and \cite{Cole94}.  Galaxy formation is modelled \emph{ab initio}, beginning with a specified cosmology and a linear power spectrum of density fluctuations and ending with predicted galaxy properties at different redshifts.  

Galaxies are assumed to form within dark matter halos, with their subsequent evolution controlled in part by the merging history of the halo.  These halo merger trees can be calculated using a Monte Carlo technique following the extended Press-Schechter formalism \citep[e.g.][]{PCH08}, or extracted directly from a dark matter only $N$-body simulation \citep[e.g.][]{Helly03,Jiang14}.  For this work we use halo merger trees derived from a Millennium-style dark matter only $N$-body simulation \citep{Springel05,Guo13}, but with cosmological parameters consistent with the 7-year \emph{Wilkinson Microwave Anisotropy Probe (WMAP7)} results \citep{Komatsu11}\footnote{$\Omega_{0}=0.272$, $\Lambda_{0}=0.728$, $h=0.704$, $\Omega_{\rm b}=0.0455$, $\sigma_{8}=0.81$, $n_{\rm s}=0.967$.}, henceforth referred to as MR7.  This simulation has a volume of ($500$ $h^{-1}$~Mpc)$^3$ and a minimum halo mass of $1.86\times10^{10}$ $h^{-1}$~M$_{\odot}$, slightly higher than the value for the original Millennium simulation ($1.72\times10^{10}$ $h^{-1}$~M$_{\odot}$).  Throughout this work we use the halo merger trees and halo masses as defined by the `Dhalo' algorithm \citep{Jiang14}.  

Some studies have shown that including  baryonic processes (e.g. AGN feedback) in $N$-body simulations can affect the matter power spectrum by $\lesssim10\%$ for scales $\lambda\lesssim5$~$h^{-1}$~Mpc at $z=0$ when compared to that of the dark matter only counterpart, due to the redistribution of gas on these scales \citep[e.g.][]{vanDaalen11}.  We note that this effect is not modelled here.  However, we are confident that our science results are robust to this as we are mostly concerned with the clustering of galaxies on larger scales.  

In \galform, the baryonic processes thought to be important for galaxy formation are included as a set of continuity equations which essentially track the exchange of mass between stellar, cold disc gas and hot halo gas components. The parameters in these equations are then calibrated against a broad range of data from both observations and simulations. Stellar population synthesis models \citep[e.g.][]{BC03,Maraston05} are used to calculate stellar luminosities.  For a more detailed description of the semi-analytic method see the reviews of \cite{Baugh06} and \cite{Benson10}.  

Various \galform models exist in the literature. For this work we use a new model (L15) which incorporates a number of important physical processes from earlier models and can reproduce an unprecedented range of observational data.  The physical processes modelled include a prescription for radio-mode AGN feedback \citep{Bower06} in which quasi-hydrostatic hot halo gas is prevented from cooling by energy input from relativistic jets, and an improved star formation law in galaxy discs based on an empirical relation between star formation rate and molecular gas \citep{BlitzRosolowsky06} first implemented in \galform by \cite{Lagos11}.  There is also a mode of star formation which takes place in a galactic bulge, triggered by either a disc instability or a galaxy merger.  Following such an event, the cold gas component in the galactic disc (formed through the cooling of hot halo gas) is transferred to a bulge/spheroid and a star formation law in which the star formation rate timescale scales with the dynamical time of the bulge is used, until this gas is exhausted.  This transfer of gas to the bulge also results in accretion onto a galaxy's central supermassive black hole (SMBH).  Throughout we use the term `starburst' to refer to a galaxy undergoing bulge star formation, and `quiescent' to mean one in which star formation occurs only in the disc.  We note that these definitions do not necessarily align with, for example, those based on a galaxy's position on the SFR-$M_{\star}$ plane.  This is discussed in more detail in a forthcoming work (Cowley et al. in preparation).

A feature of the L15 model important here is the inclusion of a top-heavy stellar initial mass function (IMF) for star formation in bursts, which allows the model to reproduce the observed number counts of galaxies selected at a range of FIR/sub-mm wavelengths \citep[$250-1100$~$\mu$m,][L15]{Cowley15} though a much less extreme IMF slope is used here than was advocated in \cite{Baugh05}\footnote{For an IMF described by ${\rm d}n/{\rm d}\ln M_{\star}\propto M_{\star}^{-x}$, $x=1$ in L15 whereas $x=0$ was used in Baugh et al.  For reference, a Salpeter (\citeyear{Salpeter55}) IMF is described by $x=1.35$.}.  A solar neighbourhood \cite{Kennicutt83} IMF is used in disc (quiescent) star formation.  

We note that we do not vary any of the fiducial L15 model parameters for this work and as such the results presented here can be considered as true predictions of the model, as it was calibrated without considering any clustering data.  

\subsection{The Dust Emission Model}
To determine a simulated galaxy's FIR/sub-mm flux, a model is required to calculate the absorption and re-emission of stellar radiation by interstellar dust.  Here, a simple model is used which assumes dust exists in two components, each with its own temperature: (i) dense molecular clouds of uniform density in which stars are assumed to form and (ii) a diffuse interstellar medium smoothly distributed throughout a double exponential disc.  

The energy of stellar radiation absorbed by each component is calculated by solving the equations of radiative transfer in this simple geometry. The dust emission is then calculated using global energy balance arguments, assuming the dust emits as a modified blackbody.  Importantly this means that the dust temperature is not a free parameter, but is calculated self-consistently for each dust component in each galaxy.  The model is therefore capable of making bona-fide multi-wavelength predictions without having to assume a shape for the spectral energy distribution (SED) of the dust emission.  

Despite its simplicity, the model is able to accurately reproduce the predictions of the more sophisticated radiative transfer code \grasil \citep{Silva98} for $\lambda_{\rm rest}\gtrsim70$~$\mu$m, thus we are confident in its application to the wavelengths under investigation in this paper.  For more details regarding the dust emission model we refer the reader to \cite{Cowley15} and the appendix of L15.    

\subsection{The Nature of Dusty Star-Forming Galaxies in the L15 model}

Here we give a brief description of the properties of the dusty star-forming galaxies which dominate the CIB and SMG population in the L15 model, in order to aid the reader in understanding results presented later.  

Dusty star-forming galaxies are predicted to be predominantly starburst galaxies (i.e. star formation occurs within the bulge), with the starburst phase being triggered by secular disc instabilities.  The importance of disc instabilities in the model is twofold: (i) they result in faster gas consumption at higher redshifts by triggering starbursts, and (ii) they are the dominant channel in the model for the growth of supermassive black holes which allow AGN feedback to suppress star formation in massive halos ($M_{\rm halo}\gtrsim10^{12}$~$h^{-1}$~M$_{\odot}$) at late times.  This means that the model displays the requisite star formation at early times to reproduce the redshift distribution of sub-millimetre galaxies at $z\gtrsim1$ without overestimating it at lower redshifts. 

Dusty star-forming galaxies are mostly central galaxies.  In the model, instantaneous ram-pressure stripping of the hot gas halo is implemented when a galaxy becomes a satellite (its hot halo gas component is transferred to that of the parent halo) and it is assumed that no more gas will accrete onto the disc of the satellite galaxy.  For this reason, the star formation in satellite galaxies is reduced due to their diminishing gas supply, and they form a minor proportion ($\lesssim5\%$) of the dusty star-forming population.

Here we present some of the physical properties of the dusty star-forming galaxy population in the L15 model, the illustrative values presented are the median values for the $L_{\rm IR}>10^{12}$~$h^{-2}$~L$_{\odot}$ population at $z=2.6$. Dusty star-forming galaxies are amongst the most massive galaxies in the simulation at a given epoch with stellar masses $M_{\star}\sim2\times10^{10}$~$h^{-1}$~M$_{\odot}$, and they reside in dark matter halos most conducive to star formation in the model ($M_{\rm halo}\sim10^{11.8}$~$h^{-1}$~M$_{\odot}$).  They also have high star formation rates $\sim140$~$h^{-1}$~M$_{\odot}$~yr$^{-1}$, translating to specific star formation rates of $\sim$~$8$~Gyr$^{-1}$ (approximately $10\times$ the sSFR of the model's `main sequence'), dust to stellar mass ratios, $M_{\rm dust}/M_{\star}\sim0.03$ and molecular gas fractions $M_{\rm cold, mol}/(M_{\rm cold, mol}+M_{\star})$~$\sim$~$0.4$. 

\section{The Spatial Clustering of Dusty Star-Forming Galaxies}

In this section we present predictions for the spatial clustering of simulated galaxies selected by their total infra-red luminosity, $L_{\rm IR}$, and their emission at $850$~$\mu$m.  We discuss how the clustering evolves with redshift, how this relates to the dark matter halos the selected objects occupy, and how the populations selected by $L_{\rm IR}$ and $S_{850\mu\rm m}$ are related.  We also briefly discuss the stellar and host halo mass of the $z=0$ descendants of the $850$~$\mu$m selected galaxies.

We present the predictions of our model in this section without considering any observational effects, such as the angular resolution of the telescopes used to identify galaxies at sub-mm wavelengths, redshift-space distortions, the accuracy of observed redshifts or any selection biases such effects can introduce.  Some of these issues are dealt with in Section \ref{sec:angular}.
\label{sec:spatial}

\subsection{The Two-Point Spatial Correlation Function}
We quantify the clustering of our selected galaxies by use of the two point spatial correlation function $\xi(r)$, which is defined as the excess probability of finding two galaxies at a given separation $r>0$, compared to a random distribution:
\begin{equation}
\delta P_{12}(r) = n^{2}[1+\xi(r)]\delta V_{1}\delta V_{2},
\label{eq:xir_def}
\end{equation}
\citep[e.g.][]{Peebles80} where $n$ is the mean number density of the selected galaxies at a given redshift and $\delta V_{\rm i}$ is a volume element.  The two-point correlation at $r=0$ is described by a Dirac delta function $\delta^{\rm D}(r)/n$ (referred to as the shot noise term) as the galaxies are treated as point objects.   

On large scales the correlation function is shaped by the clustering of galaxies in distinct dark matter halos, referred to as the two-halo term \citep[e.g.][]{CooraySheth02,BerlindWeinberg02}.  On these scales the correlation functions of the dark matter and galaxies have a similar shape but differ in amplitude.  This difference in amplitude, or bias, is defined as
\begin{equation}
b(r) = \left[\frac{\xi_{\rm gal}(r)}{\xi_{\rm DM}(r)}\right]^{1/2}\rm{ .}
\end{equation}
Although galaxy bias is scale dependent \citep[e.g.][]{Angulo08} it is usually approximated as constant on large scales, where it is governed by a weighted average of the bias values over the halos that are occupied.  The effective bias of the selected galaxy population can then be written as
\begin{equation}
b_{\rm eff} = \frac{\int b(M)n(M)\langle N_{\rm gal}|M\rangle{\rm d}M}{\int n(M)\langle N_{\rm gal}|M\rangle{\rm d}M}{\rm ,}
\label{eq:b_eff}
\end{equation}
where $b(M)$ is the bias of halos with mass $M$,  $n(M)$ is the halo mass function such that $n(M){\rm d}M$ describes the comoving number density of halos in the mass range $[M,M+{\rm d}M$], and $\langle N_{\rm gal}|M\rangle$ is the mean of the Halo Occupation Distribution (HOD, the expected number of selected galaxies within a halo of mass $M$).  

We measure the correlation function in the simulation volume using the standard estimator \citep[e.g.][]{Peebles80}:
\begin{equation}
\xi(r) = \frac{DD(r)}{N_{\rm gal}\,n\,\Delta V(r)/2}-1,
\label{eq:xir_def1}
\end{equation}
where $DD(r)$ is the number of distinct galaxy pairs with separations between $r\pm\Delta r/2$, $N_{\rm gal}$ is the total number of selected galaxies, $n$ is their mean number density and  $\Delta V(r)$ is the volume of the spherical shell between $r\pm\Delta r/2$.  We make use of the periodic nature of our simulation to calculate this volume analytically.

We calculate errors using the volume bootstrap method advocated in \cite{Norberg09}.  We divide our simulation volume into $N_{\rm sub}=27$ subvolumes and for each bootstrap realisation draw $3N_{\rm sub}$ subvolumes at random (with replacement).  As our volume is no longer periodic due to the spatial sampling we calculate $\xi(r)$ for each bootstrap realisation using the estimator presented in \cite{LandySzalay93}:
\begin{equation}
\xi(r) = \frac{DD(r)-2DR(r) +RR(r)}{RR(r)}{\rm ,}
\label{eq:LandySzalay}
\end{equation} 
where $DD(r)$, $DR(r)$ and $RR(r)$ represent the number of data-data, data-random, and random-random pairs with separations between $r\pm\Delta r/2$.  For each bootstrap realisation we generate a random catalogue with $10$ times more points than there are galaxies in our initial sample, normalising the $DR$ and $RR$ terms in equation (\ref{eq:LandySzalay}) to have the same total number of pairs as $DD$.  We calculate $100$ bootstrap realisations from which we derive the $1\sigma$ percentile variation for each bin of separation.
    
\subsection{Spatial Clustering Evolution of Infra-red Luminous Galaxies}
\begin{figure*}
\centering
\includegraphics[width=\linewidth]{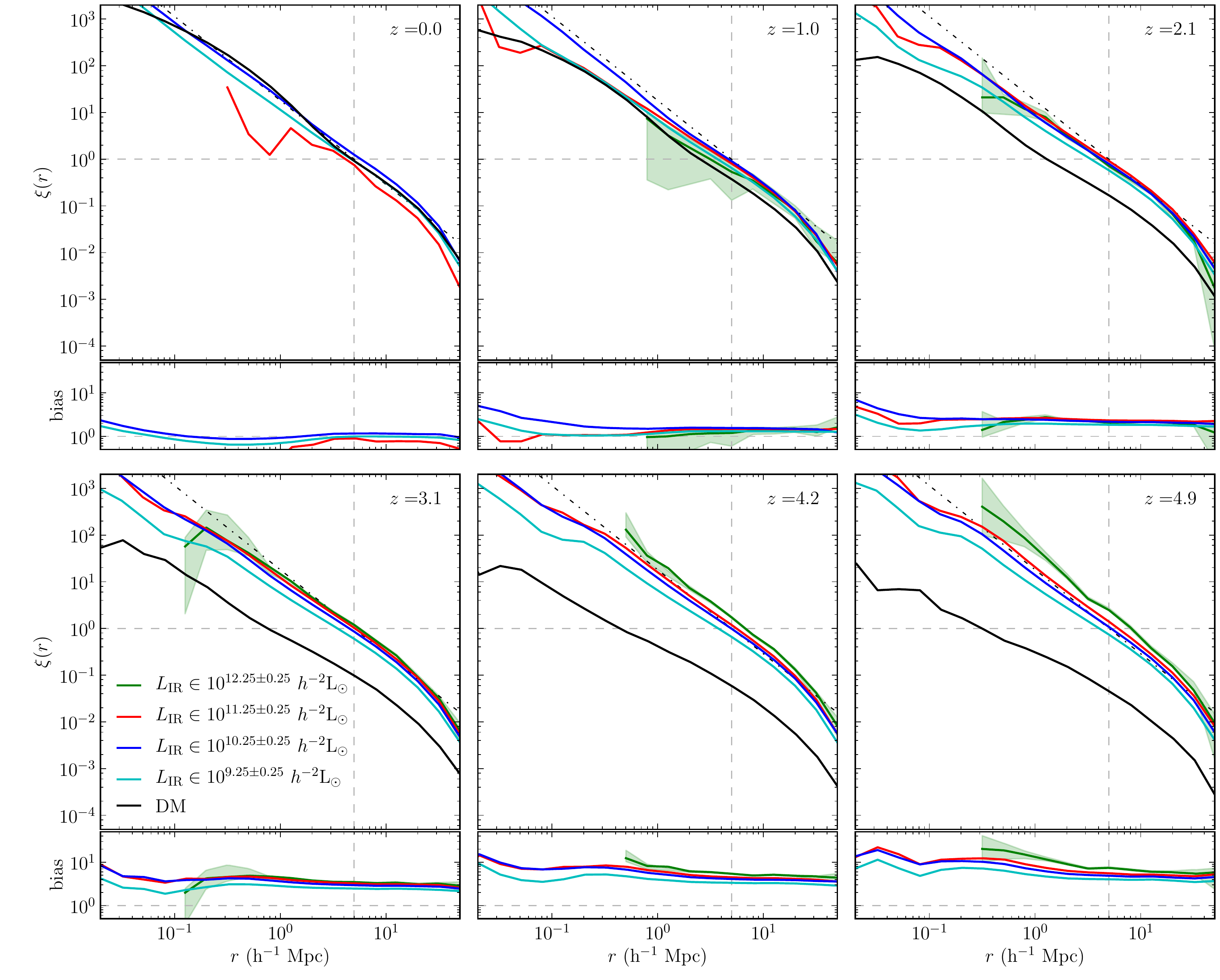} 
\caption{\emph{Main panels:}  The predicted two-point spatial correlation function, $\xi(r)$, as a function of comoving separation, $r$, for galaxies selected by their total $8-1000$~$\mu$m luminosity, $L_{\mathrm{IR}}$, at the redshift indicated in each panel.  The cyan, blue, red and green lines show galaxies with $L_{\rm IR}=10^{9}-10^{9.5}$, $10^{10}-10^{10.5}$, $10^{11}-10^{11.5}$ and $10^{12}-10^{12.5}$~$h^{-2}$~L$_{\odot}$ respectively.  The green shaded region shows the $1\sigma$ volume bootstrap errors for the $L_{\rm IR}=10^{12}-10^{12.5}$~$h^{-2}$~L$_{\odot}$  population.  The black line indicates the correlation function measured for dark matter particles in the MR7 simulation.  The vertical and horizontal dashed grey lines are drawn for reference at $r=5$ $h^{-1}$~Mpc and $\xi=1$ respectively.  The diagonal black dash-dotted line, again for reference, indicates a $\gamma = -1.8$ power law with a correlation length of $5$~$h^{-1}$~Mpc.  \emph{Sub panels:} As for the main panels but indicating the bias, defined as $(\xi_{\rm g}/\xi_{\rm DM})^{1/2}$.  A horizontal grey dashed line indicating a bias value of $1$ is drawn for reference in each panel.}
\label{fig:Ldsol_xir_bias}
\end{figure*}
\begin{figure*}
\centering
\includegraphics[width=\linewidth]{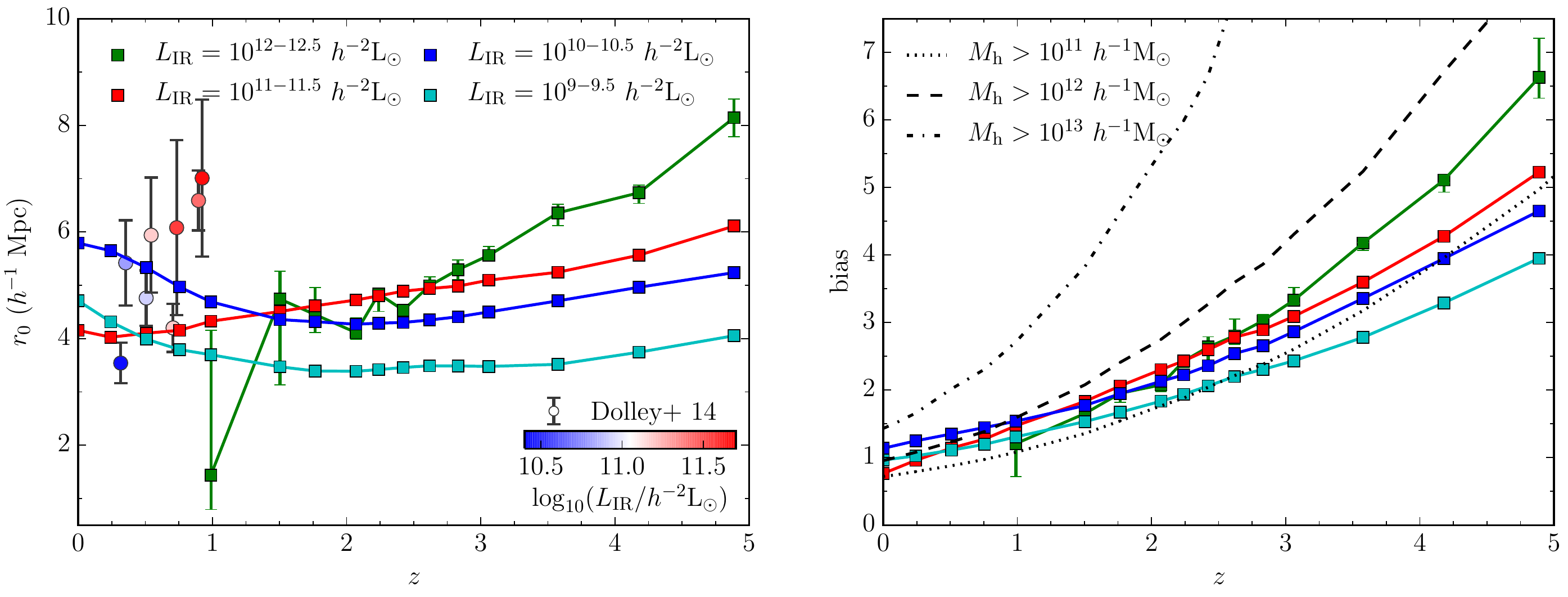}
\caption{\emph{Left panel:} Evolution of the comoving correlation length $r_{0}$ [defined such that $\xi(r_{0})\equiv1$].  The cyan, blue, red and green lines show galaxies with $L_{\rm IR}=10^{9}-10^{9.5}$, $10^{10}-10^{10.5}$, $10^{11}-10^{11.5}$ and $10^{12}-10^{12.5}$~$h^{-2}$~L$_{\odot}$ respectively.  The errors indicate $1\sigma$ volume bootstrap errors for the $L_{\rm IR}=10^{12}-10^{12.5}$~$h^{-2}$~L$_{\odot}$ population.  A selection of observational estimates from Dolley et al. (\citeyear{Dolley14}) are shown as circles, with the colour scale indicating the mean $L_{\rm IR}$ for each sample, as shown on the inset colour bar.  \emph{Right panel:} As for the left panel, but indicating the evolution of the large scale bias.  The dotted, dashed and dash-dotted lines indicate the bias evolution for halos of $M_{\rm h}>$ $10^{11}$, $10^{12}$ and $10^{13}$ $h^{-1}$~M$_{\odot}$ respectively, as measured directly from the MR7 simulation.}
\label{fig:Ldsol_r0_evolution}
\end{figure*} 
\begin{figure*}
\centering
\includegraphics[width=\linewidth]{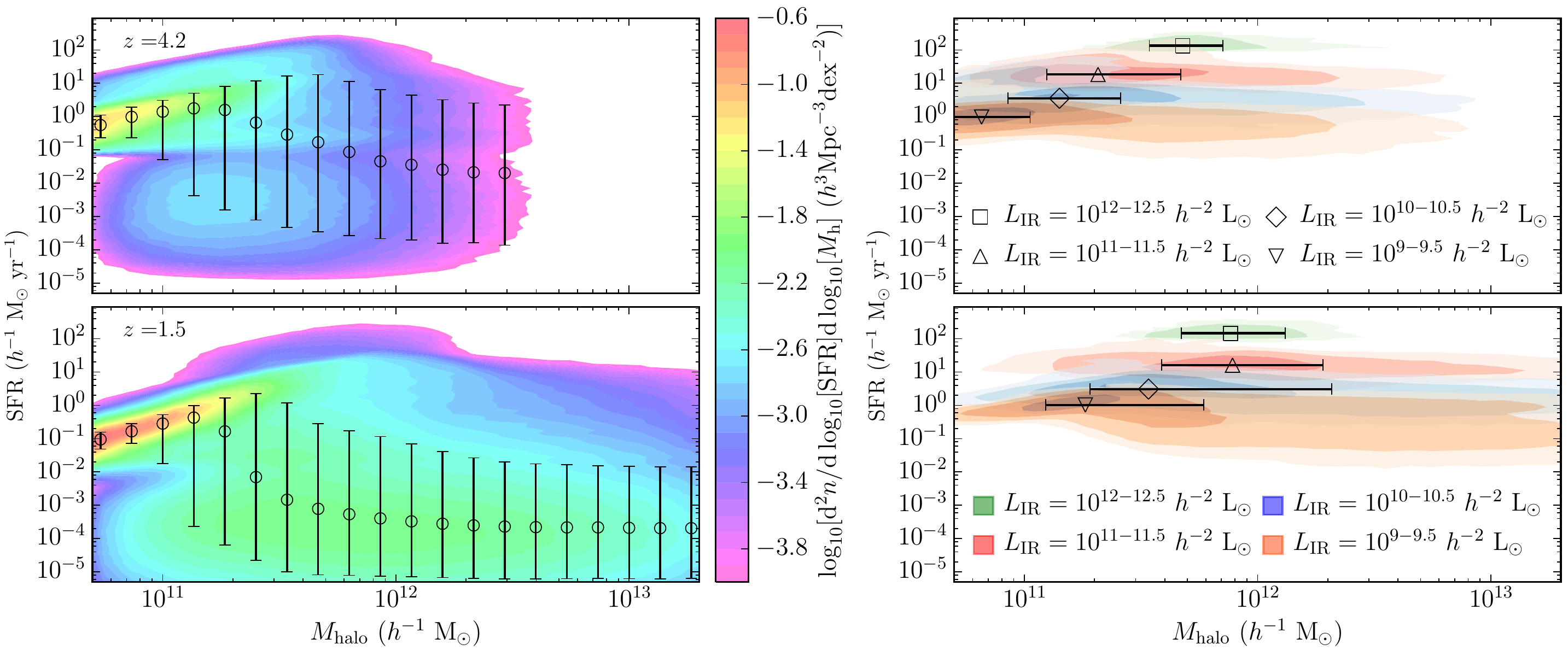}
\caption{Predicted distribution of galaxies in the star formation rate - halo mass plane at $z=4.2$ (top panels) and $z=1.5$ (bottom panels).  \emph{Left panels:} Distribution of all galaxies with the shading representing the galaxy number density at that position on the plane, with red indicating the highest number densities and purple the lowest.  Open circles show the median SFR in bins of halo mass, with the errorbars indicating the $16-84^{\mathrm{th}}$ percentile scatter.  \emph{Right panels:} Distribution of galaxies selected by their total infra-red luminosity for luminosities of $10^{12}-10^{12.5}$ (green), $10^{11}-10^{11.5}$ (red), $10^{10}-10^{10.5}$ (blue) and $10^{9}-10^{9.5}$~$h^{-2}$~L$_{\odot}$ (orange contours).  The open symbols indicate the median halo mass and SFR in the corresponding luminosity bin, with the errorbars indicating the $16-84^{\mathrm{th}}$ percentile scatter in halo mass.}
\label{fig:SFR_mhalo}
\end{figure*} 

Here we present predictions for the clustering of galaxies selected by their total infra-red luminosity $L_{\rm IR}$,  derived by calculating the energy of stellar radiation absorbed by dust through solving the equations of radiative transfer in our assumed dust geometry.

We show the model predicted spatial clustering for galaxies selected by their $L_{\rm IR}$ in Fig. \ref{fig:Ldsol_xir_bias} at a selection of redshifts, $z\sim0-5$, and luminosities, $L_{\rm IR}\sim10^{9}-10^{12.5}$~$h^{-2}$~L$_{\odot}$.  For clarity, we only show volume bootstrap errors for the most luminous (i.e. least numerous) population.  We are confident that our selected galaxies are complete populations, at all redshifts considered here, and are not affected by the finite halo mass resolution of our simulation.  We also plot the correlation function of the dark matter, calculated using a randomly chosen subset of $10^6$ dark matter particles from the MR7 simulation, and can see that the selected galaxy populations represent biased tracers of the underlying matter distribution.  Note that we do not show $\xi(r)$ of the most luminous population at $z<1$ as the number of pairs of such objects in our simulation at these redshifts is not sufficient to provide a robust prediction.

It is notable that the clustering of the selected galaxies shows a dependence on the selection luminosity, and redshift.  This is summarised in Fig.~\ref{fig:Ldsol_r0_evolution}, which shows the redshift evolution of the comoving correlation length, $r_{0}$, defined such that $\xi(r_{0})\equiv1$, and the large scale bias of the selected populations.  In the right panel of Fig.~\ref{fig:Ldsol_r0_evolution} we show for reference the large-scale bias evolution of halos selected by their mass, calculated directly from the MR7 simulation.

At all redshifts shown the two fainter luminosity populations are predominantly composed of quiescently star-forming galaxies, they display a similar clustering evolution, though systematically offset such that the brighter of these two populations is more clustered at all redshifts.  The brighter two populations are predominantly composed of starburst galaxies\footnote{The luminosity at which the infra-red luminosity functions predicted by our model become dominated by starburst galaxies is dependent on redshift.  For example, at $z=0$ the luminosity function is dominated by starbursts for $L_{\rm IR}\gtrsim10^{11.3}$~$h^{-2}$~L$_{\odot}$, at $z=4.9$ this limit is  $L_{\rm IR}\gtrsim10^{10.5}$~$h^{-2}$~L$_{\odot}$.} and display a different clustering evolution to the fainter two samples, with $r_{0}$ displaying a monotonic relationship with redshift.

Comparing with the large-scale bias evolution of mass-selected halos we can see that our most luminous population displays an evolution consistent with them residing in halos of mass $10^{11}-10^{12}$~$h^{-1}$~M$_{\odot}$ over the redshift range $z\sim1-5$.  

These results can be understood better in the context of the halo masses sampled by the infra-red luminosity selection.  In Fig.~\ref{fig:SFR_mhalo} we show the distribution of galaxies in the star formation rate - halo mass plane for all galaxies (left panels) and for the infra-red luminosity selected populations (right panels).  We can see that the distribution of SFRs is broad for halo masses $M_{\rm halo}>10^{11}$~$h^{-1}$~M$_{\odot}$ and that the infra-red selections pick up a broad range of halo masses.  We also see how this distribution evolves.  At $z=4.2$ the infra-red selection means that samples with increasing $L_{\rm IR}$ have increasing median halo masses, leading to them being more biased than samples selected by a lower infra-red luminosity.  At $z=1.5$ this is no longer the case, as the most luminous population has a slightly lower median halo mass than the next most luminous.  This breaks the monotonic relation of increasing bias with increasing luminosity seen at higher redshifts.   

In Fig.~\ref{fig:Ldsol_r0_evolution} we also compare our predictions to the observational estimates of \cite{Dolley14}, who used far infra-red luminosities derived from $24$~$\mu$m fluxes.  We show the $r_{0}$ values for their redshift bins that are complete in infra-red luminosity, for clarity showing only most and least luminous samples within each redshift bin.  The colour scale indicates the mean infra-red luminosities of their samples, the bins for which have a width of 0.25~dex in $L_{\rm IR}$.  Whilst the overall agreement is generally favourable, Dolley et al. find, in contrast to our predictions, that for $z<1$ at a fixed redshift $r_{0}$ increases with increasing luminosity.  The model also appears to underpredict the clustering of $\sim10^{11.5}$~$h^{-2}$~L$_{\odot}$ galaxies at $z\sim1$ and overpredict the clustering of $\sim10^{10.5}$~$h^{-2}$~L$_{\odot}$ galaxies at $z\sim0.3$.  

There could be a number of reasons for this discrepancy. Dolley et al. assumed a power-law slope of $\gamma=1.9$ in order to derive a correlation length. If a lower value is used (as favoured by our model) they note that this increases their estimated correlation lengths (e.g. assuming $\gamma=1.8$ gave correlation lengths $\sim0.5$~$h^{-1}$~Mpc larger).  Our model shows a variation of power-law slope with redshift and infra-red luminosity, with lower luminosity samples having generally flatter slopes.  It is also unclear whether the simulated galaxies follow the relation used by Dolley et al. to derive $L_{\rm IR}$ from the observed $24$~$\mu$m photometry, which is based on templates derived from local galaxies \citep{Rieke09} and adjusted at higher redshifts according to \cite{Rujopakarn13}.  Alternatively, further investigation into the physical processes which produce the distribution of galaxies on the SFR-$M_{\rm halo}$ plane as predicted by the model (Fig.~\ref{fig:SFR_mhalo}) is required to understand how the predicted clustering could be brought into better agreement with the Dolley et al. results.            

Our predictions for correlation length in this section are lower than the observational estimates of \cite{Farrah06}, who infer correlation lengths of $9.4\pm2.2$ and $14.4\pm2.0$~$h^{-1}$~Mpc for galaxies at $z \sim1.7$ and $2.5$ respectively, with $L_{\rm IR}\gtrsim 5\times 10^{11}$~$h^{-2}$~L$_{\odot}$.  However, we do not consider this a significant discrepancy, due to the complicated selection criteria of the Farrah et al. sample, which we do not attempt to model here, and assumptions made by those authors regarding the redshift distribution of their sample, and their parametrisation of $\xi(r,z)$.

\subsection{Spatial Clustering of SMGs}
\label{subsec:spatial_SMGs}
\begin{figure*}
\centering
\includegraphics[width=0.95\linewidth]{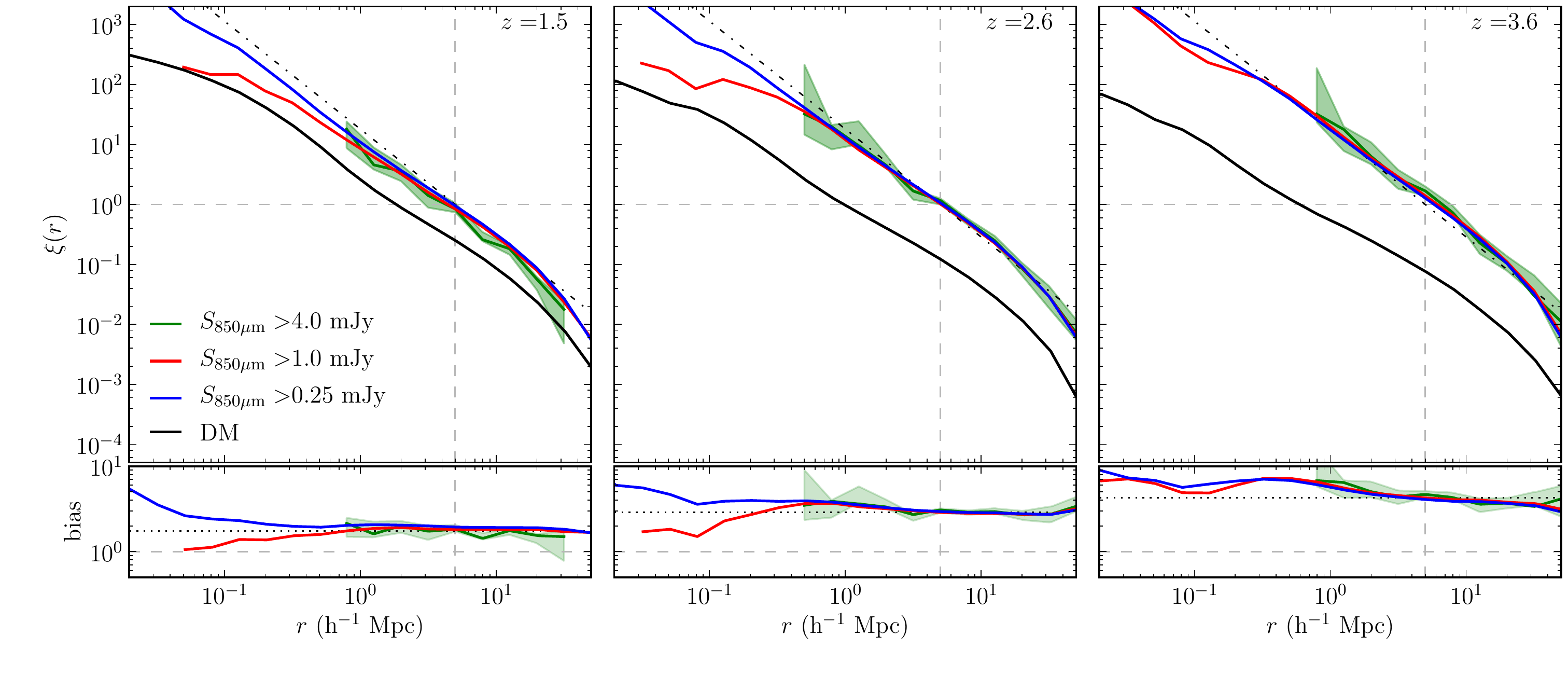}
\caption{\emph{Top panels:} The spatial correlation function for $850$~$\mu$m selected galaxies at redshifts of $1.5$, $2.5$ and $3.5$ (left to right).  The blue, red and green lines show the correlation function for $S_{850 \mu\rm m} >$ $0.25$, $1.0$ and $4.0$~mJy respectively.  The green shaded region shows the $1\sigma$ volume bootstrap errors for the $S_{850\mu\rm m}>4.0$~mJy population.  The black line indicates the correlation function measured for dark matter particles in the MR7 simulation.  The vertical and horizontal dashed grey lines are drawn for reference at $r=5$ $h^{-1}$~Mpc and $\xi=1$ respectively.  The diagonal black dash-dotted line, again for reference, indicates a $\gamma = -1.8$ power law with a correlation length of $5$ $h^{-1}$~Mpc.  \emph{Bottom panels:} As for the top panel but indicating the bias, defined as $(\xi_{\rm g}/\xi_{\rm DM})^{1/2}$.  A horizontal grey dashed line, drawn for reference in each panel, indicates a bias of $1$.  A horizontal black dotted line, again drawn for reference, indicates a bias of $1.7$, $2.9$ and $4.2$ (left to right). }
\label{fig:xir_bias}
\end{figure*}

In this section we present the spatial clustering of galaxies selected by their $850$~$\mu$m flux.  We focus on this wavelength as it is historically the wavelength at which the majority of ground-based observations of FIR/sub-mm galaxies have been performed, due to the atmospheric transmission window.  The real space two-point correlation function and large-scale bias for our selected galaxies are presented in Fig. \ref{fig:xir_bias} 
 over a range of redshifts which span the peak of the redshift distribution of the selected SMGs.  

We consider three samples of galaxies selected by flux: (i) a bright population with $S_{850\mu \rm m}>4$~mJy (median $L_{\rm IR}\sim 10^{12.2}$~$h^{-2}$~L$_{\odot}$ at $z=2.6$, green line) as this is a typical limit at which single-dish surveys can detect SMGs \citep[e.g.][though note we do not consider the effects of the single-dish beam in this section]{Weiss09}, (ii) an intermediate population with $S_{850\mu\rm m}>1$~mJy (median $L_{\rm IR}\sim 10^{11.8}$~$h^{-2}$~L$_{\odot}$ at $z=2.6$, red line) as this is an approximate limit to which ALMA detected galaxies as part of Cycle 0 observations \citep[e.g.][]{Hodge13} and (iii) a faint population with $S_{850\mu\rm m}>0.25$~mJy (median $L_{\rm IR}\sim 10^{11.2}$~$h^{-2}$~L$_{\odot}$ at $z=2.6$, blue line) which are in principle detectable by ALMA, though with longer integration times and more antennae than were used in Cycle 0.  Our selected galaxies exhibit clustering with $r_{0}\sim5$~$h^{-1}$~Mpc, with little dependence on flux, for the fluxes considered here.      

\subsubsection{SMG Halo Occupation Distribution}
\begin{figure*}
\includegraphics[width=0.8\linewidth]{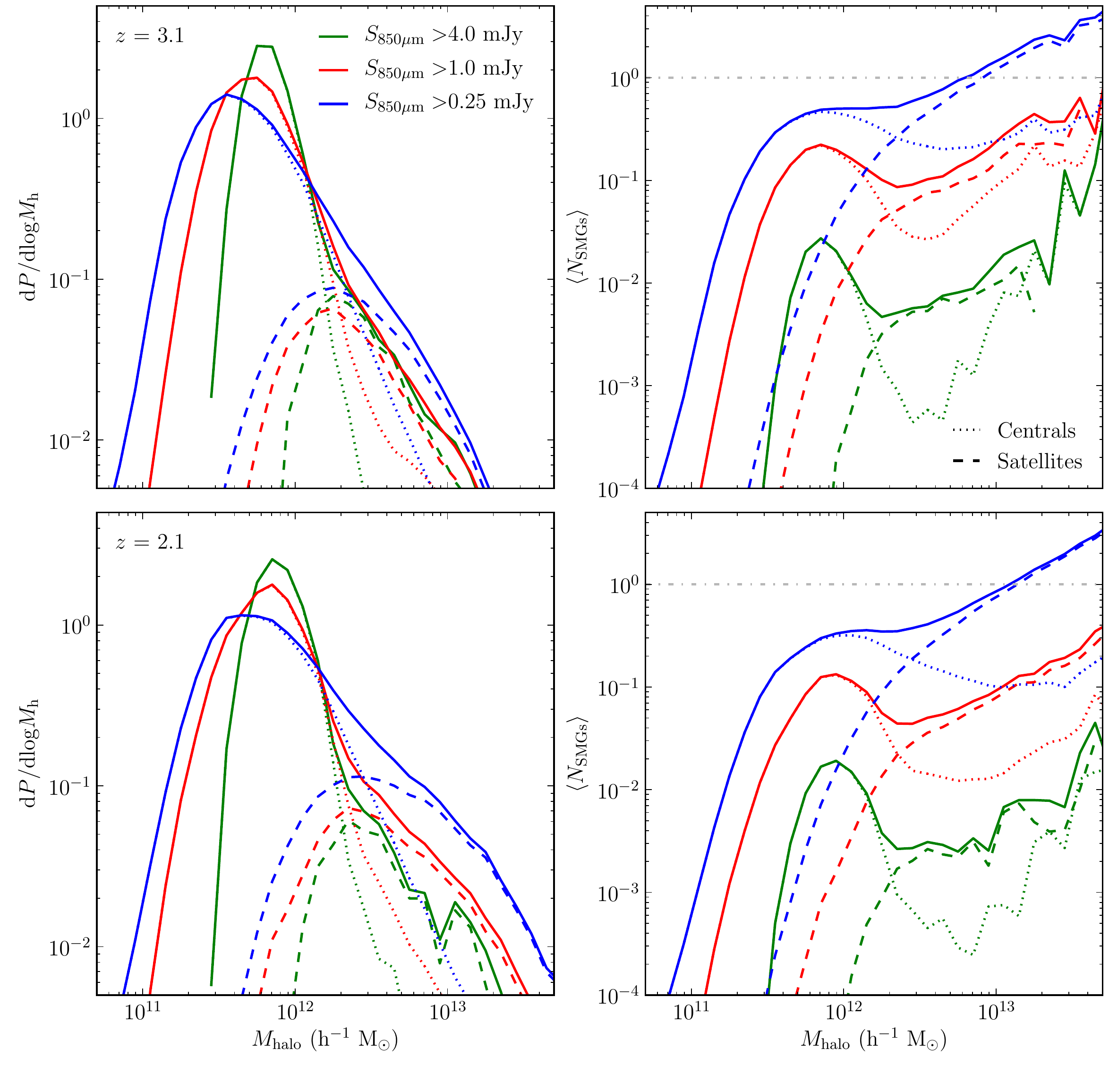}
\caption{Probability distribution of halo mass (left) and halo occupation distribution (right) for $850$~$\mu$m selected SMGs at $z=3.1$ (top) and $2.1$ (bottom).  The blue, red and green lines indicate the HOD for the $S_{850\mu\rm{m}}>0.25$, $1.0$ and $4.0$~mJy respectively, with the dashed (dotted) lines depicting satellite (central) galaxies.  A horizontal dash-dotted line is drawn in both right panels at $\langle N_{\rm{SMGs}}\rangle=1$ for reference.}
\label{fig:HOD}
\end{figure*}

We can gain further insight into the clustering of the selected SMGs from Fig. \ref{fig:HOD} which shows their halo mass probability distribution (i.e. the product of the halo mass function and the mean of the HOD - $n(m)\langle N_{\rm gal}|M\rangle$ in equation \ref{eq:b_eff}, left panels) and the mean of the HOD ($\langle N_{\rm gal}|M\rangle$ in equation \ref{eq:b_eff}, right panels) at redshifts $z=3.1$ and $2.1$ (top and bottom panels respectively).  
It is evident from the left panels that SMGs reside predominantly in halos of mass $\sim10^{11.5}-10^{12}$~$h^{-1}$~M$_{\odot}$, the halo mass range most conducive for star formation in our model over a broad range of redshifts (see Fig. 27 of L15).  For example, at $z=3.1$: $87$, $74$ and $54$\% of galaxies in the $S_{850\mu\rm m}>4$, $1$ and $0.25$ mJy selected populations respectively reside in halos within this mass range.  At $z=2.1$ these percentages are $75$, $69$ and $53$\% respectively.  The halo mass at which the probability distribution peaks seems insensitive to the $850$~$\mu$m flux of the galaxies and their redshift,  although fainter galaxies do occupy a broader range of halo masses, and the distribution for satellite galaxies (dashed lines) peaks at a higher halo mass ($\sim 2\times10^{12}$~$h^{-1}$~M$_{\odot}$). 

In the right panels the HODs for central galaxies (dotted lines) peak below unity for all samples.  The HODs only reach unity for satellites in fainter samples in massive halos ($M_{\rm h}\gtrsim10^{13}$ $h^{-1}$~M$_{\odot}$ at $z=2.1$).  Models which force $\langle N_{\rm SMGs,c}\rangle=1$ and adopt the same number density of SMGs would place them in more massive halos than predicted by our model.  An $S_{850 \mu\rm m}>1$~mJy galaxy is hosted in roughly 1 in every 10 halos of $\sim10^{12}$ $h^{-1}$~M$_{\odot}$, showing the need for a large number of halo histories to be sampled (i.e. large cosmological volumes simulated) in order to make robust predictions for the SMG population as a whole \citep[see also e.g.][]{Almeida11,Miller15}).  

We attribute the minima in the HODs for the central galaxies to merger-induced SMGs.  In our model AGN feedback becomes effective in massive haloes ($M_{\rm halo}\gtrsim10^{12}$~$h^{-1}$~M$_{\odot}$), which prevents hot halo gas from cooling, limiting the fuel for star formation and leading to the downturn in the HOD.  Galaxy mergers bring in a fresh reservoir of cold gas to central galaxies, allowing further star formation in these high mass ($\gtrsim10^{13}$~$h^{-1}$~M$_{\odot}$) halos without the need for in-situ gas cooling.   

\subsubsection{The Evolution of SMG Clustering}
\begin{figure*}
\includegraphics[width=\linewidth]{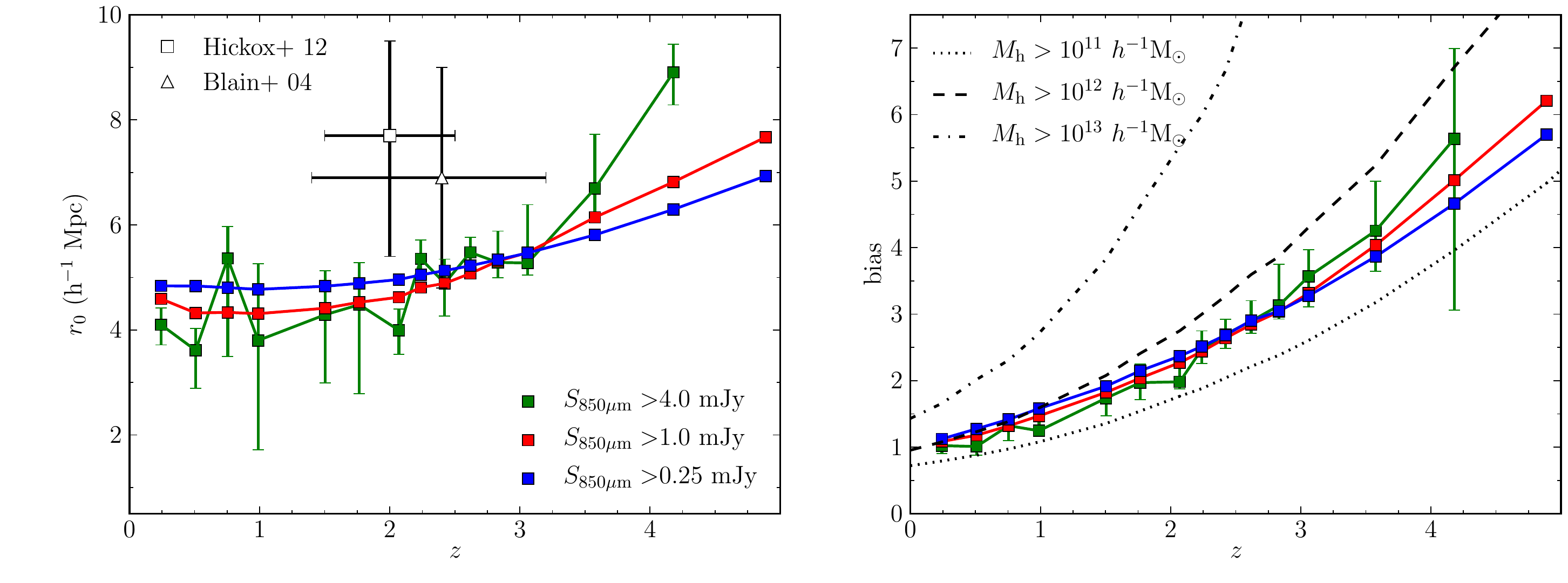}
\caption{\emph{Left panel:} Evolution of the comoving correlation length $r_{0}$ [defined such that $\xi(r_{0})\equiv1$] with redshift, for galaxies with $S_{850\mu\rm m}>0.25$, $1.0$ and $4.0$  mJy (blue, red and green lines respectively).  The errors indicate $1\sigma$ volume bootstrap errors for the $S_{850\mu\rm m}>4.0$~mJy population.  The observational data are taken from Hickox et al. (\citeyear{Hickox12}; squares) and Blain et al. (\citeyear{Blain04}; triangles).  \emph{Right panel:} Symbols and coloured lines as for the left panel but indicating the evolution of the large scale bias.  The dotted, dashed and dash-dotted lines indicate the bias evolution for halos of $M_{\rm halo}>$ $10^{11}$, $10^{12}$ and $10^{13}$ $h^{-1}$~M$_{\odot}$ respectively, as measured directly from the MR7 simulation.}
\label{fig:r0_evolution}
\end{figure*}  
We show the evolution of the correlation length $r_{0}$ in the left panel of Fig. \ref{fig:r0_evolution}.  This is approximately constant for $z\lesssim2$ but increases with increasing redshift at higher redshifts.  The errorbars shown are derived from the $1\sigma$ bootstrap errors described above. 

In the right panel of Fig. \ref{fig:r0_evolution} we show the evolution of the large scale bias with redshift, in addition plotting for reference the evolution of the large-scale bias for halos selected by their mass.  We can see that the bias evolution of our galaxies is of a similar form to that of the halos, indicating that SMGs typically reside in halos of $10^{11}-10^{12}$~$h^{-1}$~M$_{\odot}$, over a large redshift range.  This is in agreement with our previous findings in Fig.~\ref{fig:HOD}.  

In Fig.~\ref{fig:r0_evolution} we compare to the observational results of \cite{Hickox12} and \cite{Blain04}.  Hickox et al. use sub-mm sources from the single-dish LESS source catalogue \citep[][]{Weiss09}, with $S_{850\mu\rm m}\gtrsim4.5$~mJy, at redshifts of $z\sim2-4$, covering $0.35$~deg$^{2}$, and use the cross-correlation of these with IRAC selected galaxies over a similar redshift range,  taking into account the photometric redshift probability distribution of their SMGs \citep{Wardlow11}, to derive a large scale bias of $3.4\pm0.8$  from which they find a correlation length of $r_{0}=7.7_{-2.3}^{+1.8}$~$h^{-1}$~Mpc assuming a power-law correlation function [$\xi(r)=(r/r_{0})^{-\gamma}$] with $\gamma = 1.8$.  Blain et al. also assume a power-law $\xi(r)$ with $\gamma = 1.8$, and a Gaussian redshift distribution \citep{Chapman05}, whilst allowing $r_{0}$ to vary in order to match the number of SMG ($S_{850\mu\rm m}\gtrsim5$~mJy) pairs observed across a number of non-contiguous SCUBA fields with a combined area of $\sim 0.16$~deg$^{2}$. They obtain a correlation length of $r_{0}=6.9\pm2.1$ $h^{-1}$~Mpc but note that if they exclude the most overdense field from their analysis, they derive $r_{0}=5.5\pm1.8$ $h^{-1}$~Mpc, which is in better agreement with our predictions.  However, due to the significant errors on the observational data and potential biases due to the single-dish beam used in these studies which we discuss in Section~\ref{sec:angular}, it is difficult to draw any strong conclusions about the level of agreement between the model and data.

From comparing the left panel of Fig. \ref{fig:r0_evolution} to that of Fig. \ref{fig:Ldsol_r0_evolution}, we can see that the clustering evolution of our SMG populations are remarkably similar to that of our most infra-red luminous galaxies ($L_{\rm IR}=10^{12}-10^{12.5}$~$h^{-2}$~L$_{\odot}$). We note that at $z=2.6$ the median $850$~$\mu$m flux for galaxies in our most luminous $L_{\rm IR}$ bin ($10^{12}-10^{12.5}$~$h^{-2}$~L$_{\odot}$) is $3.3^{+2.2}_{-1.5}$ mJy, where the errorbars represent the $10-90$ percentiles.  Conversely, at the same redshift the $S_{850\mu\rm m}>4$~mJy population has a bolometric dust luminosity of $L_{\rm IR}=10^{12.04}-10^{12.44}$~$h^{-2}$~L$_{\odot}$ (10-90 percentiles).  Thus in our model the $850$~$\mu$m selection selects the most infra-red luminous starburst galaxies (our predicted galaxy number counts at $850$~$\mu$m are dominated by starburst galaxies for $S_{850\mu\rm m}\gtrsim0.2$~mJy), hence the similarities in the model predicted clustering evolution of SMGs and the most infra-red luminous galaxies.

\subsubsection{SMG Descendants and Environment}
\begin{figure}
\includegraphics[width=\columnwidth]{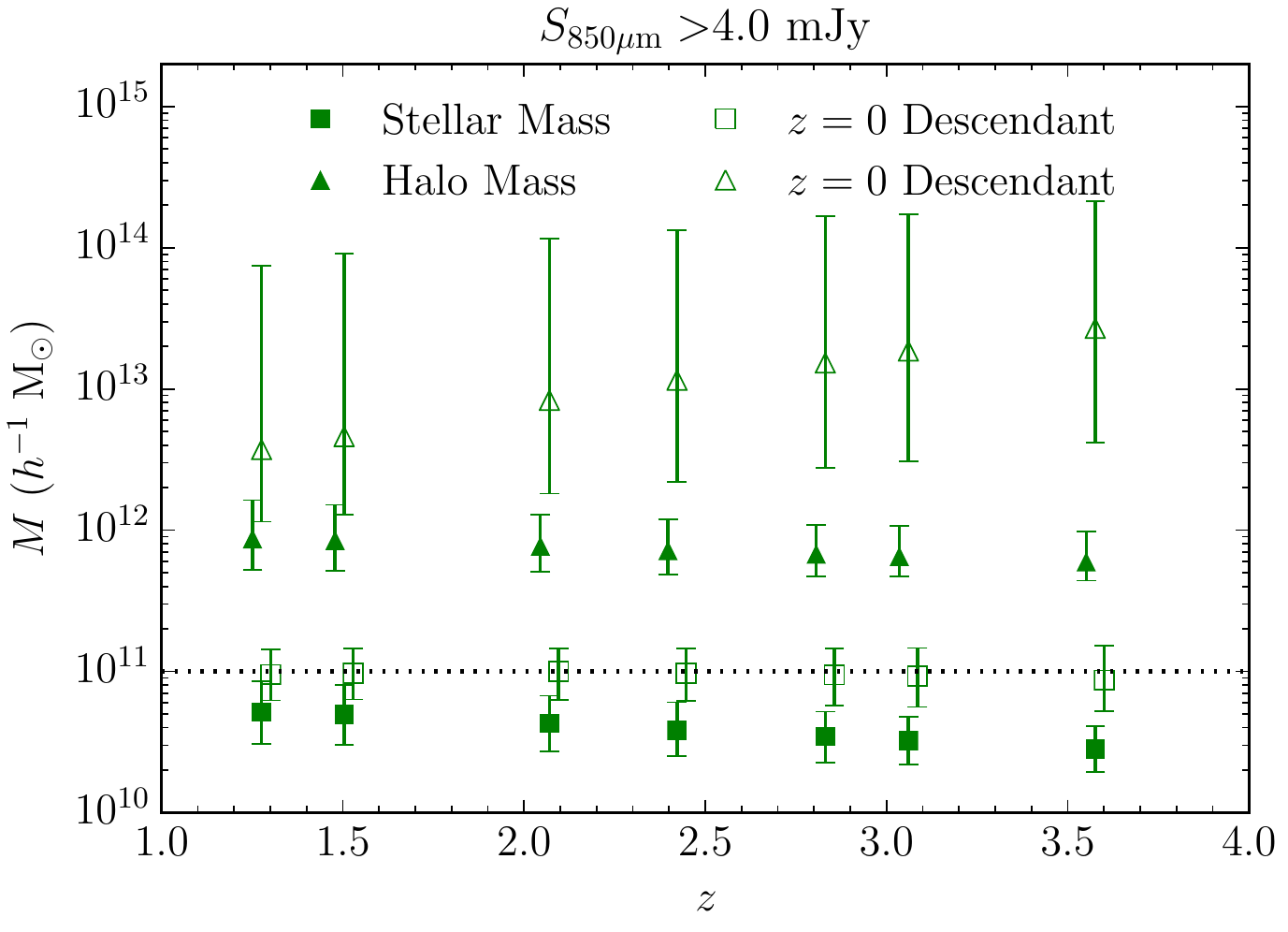}
\caption{The descendants of $S_{850\mu\rm m}>4.0$~mJy selected galaxies in our simulation.  The squares and triangles indicate the median stellar and host halo mass of the selected galaxies respectively, with the filled symbols indicating this quantity at the redshift of interest and the open symbols indicating this quantity for the $z=0$ descendant.  The error bars indicate $10-90$ percentile ranges.  The open squares and filled triangles are offset in redshift by $\pm0.025$ for clarity.  A dotted horizontal line is drawn at $M=10^{11}$~$h^{-1}$~M$_{\odot}$ for reference.} 
\label{fig:SMG_descendants}
\end{figure}

Arguments which assume that the majority of $z=0$ stellar mass of an SMG descendant is formed during the sub-mm bright phase imply that by fading the stellar population, SMGs could evolve onto the $z=0$ scaling relations of massive ellipticals \citep[assuming a burst duration of typically $\sim100$~Myr, e.g.][]{Swinbank06,Simpson13}.  Here we investigate the stellar and halo masses of the $z=0$ descendants, presenting our findings for the bright population ($S_{850\mu\rm m}>4$~mJy) in Fig. \ref{fig:SMG_descendants}.

We find that across all redshifts shown in Fig. \ref{fig:SMG_descendants}, which span the majority of the redshift distribution for this population, the selected galaxies evolve into galaxies with a stellar mass of $\sim10^{11}$~$h^{-1}$~M$_{\odot}$ at the present day.  This is similar to the results presented from an analysis of an earlier version of the galaxy formation model used here \citep{Gonzalez11}.  

The stellar masses of SMGs inferred from observations are the subject of much debate.  They are typically inferred by SED fitting to broadband photometry, making a range of assumptions regarding the AGN contamination, dust obscuration, star formation history and IMF of the galaxies in question.  Early estimates appeared to disagree by factors of $\sim5-10$ for the \emph{same} sample of SMGs. \cite{Hainline11} quoted a median stellar mass for the \cite{Chapman05} sample ($S_{850\mu\rm m}>5$~mJy) of $\sim5\times10^{10}$~$h^{-1}$~M$_{\odot}$ [assuming a \cite{Kroupa02} IMF] in contrast to the higher value of $\sim2.6\times10^{11}$~$h^{-1}$~M$_{\odot}$ found by \cite{Michalowski10} [assuming a \cite{Chabrier03} IMF], though subsequent work by \cite{Michalowski12b} suggested that this discrepancy was mostly due to the assumed star formation histories used by the two studies, once differences due to the choice of IMF were taken into account.  \cite{Michalowski12b} also obtained a revised median stellar mass of $\sim1.4\times10^{11}$~$h^{-1}$~M$_{\odot}$.  More recently \cite{daCunha15} derive a median stellar mass of $\sim6\times10^{10}$~$h^{-1}$~M$_{\odot}$ by applying the SED fitting code \magphys [assuming a \cite{Chabrier03} IMF] to the ALESS \citep{Hodge13} SMG sample.
 
Our predicted stellar masses lie at the lower end of values quoted in the literature however, it is difficult to understand the significance of the (dis)agreement. The comparison is further complicated by the top-heavy IMF for starbursts assumed in the model.  We therefore consider a proper comparison of the stellar masses of SMGs predicted by our model and those inferred from observations to be beyond the scope of this paper, and caution the reader against over-interpreting the values presented briefly here. 

The predicted masses presented in Fig.~\ref{fig:SMG_descendants} are qualitatively similar for the fainter populations, though they systematically shift to slightly lower masses, for example the $S_{850\mu\rm m}>0.25$~mJy population evolve into galaxies with stellar mass $\sim5\times10^{10}$~$h^{-1}$~M$_{\odot}$. Note also that here we consider unique descendants, such that if two galaxies selected at a given redshift evolve into the same descendant at $z=0$ it is only counted once.

In terms of halo mass, whilst sub-mm selected galaxies occupy a relatively narrow range of halo masses ($\sim0.5$~dex) at the redshift at which they are selected, the range of halo masses which host the $z=0$ descendants is broad, spanning nearly two orders of magnitude $\sim10^{12}-10^{14}$~$h^{-1}$~M$_{\odot}$.  In our model it appears then that bright SMGs do not necessarily trace the most massive $z=0$ environments.  As with stellar mass, here we consider unique halos, such that if a halo contains two galaxies selected at a given redshift, or the $z=0$ descendant(s) of two galaxies selected at a given redshift, it is only counted once.

Our results for stellar and halo masses of bright SMGs and their descendants are a factor of $\sim5$ lower than those found by \cite{Munoz15}. However, their simulations do not self-consistently predict the sub-mm flux of galaxies as is done in this work, but instead rely on a `count-matching' approach to link a galaxy's physical properties to its sub-mm flux. They infer median stellar and halo mass of $10^{11.2}$ and $10^{12.7}$~$h^{-1}$~M$_{\odot}$ respectively for SMGs; and $10^{11.7}$ and $10^{13.8}$~$h^{-1}$~M$_{\odot}$ respectively for the $z=0$ descendants of SMGs. 

\section{Angular clustering at 850 $\mu$\lowercase{m}}
\label{sec:angular}
The simplest measure of clustering from a galaxy imaging survey is the angular two-point correlation function $w(\theta)$.  Analogously to equation (\ref{eq:xir_def}), the probability of finding two objects separated by an angle $\theta>0$\footnote{Analogously to the spatial case, at $\theta=0$ the correlation function is described by a Dirac delta function, $\delta^{\rm D}(\theta)/\eta$, which is referred to as the shot noise term.} is defined as:
\begin{equation}
\delta P_{12}(\theta) = \eta^{2}[1+w(\theta)]\delta\Omega_{1}\delta\Omega_{2},
\end{equation}
where $\eta$ is the mean surface density of objects per unit solid angle and $\delta\Omega_{\rm i}$ is a solid angle element, such that $w(\theta)$ represents the excess probability of finding objects at angular separation $\theta$, compared to a random (Poisson) distribution.

In this section we present the angular correlation function of galaxies, $w_{\rm g}$, selected by their $850$~$\mu$m emission.  We compare this to the angular correlation function of sub-mm sources, $w_{\rm s}$, extracted from simulated single-dish $850$~$\mu$m imaging following the method presented in \cite{Cowley15}, and the angular correlation function of $850$~$\mu$m intensity fluctuations, $w_{\rm I}$.

\subsection{The Angular Clustering of Galaxies}

\begin{figure}
\includegraphics[width=\columnwidth]{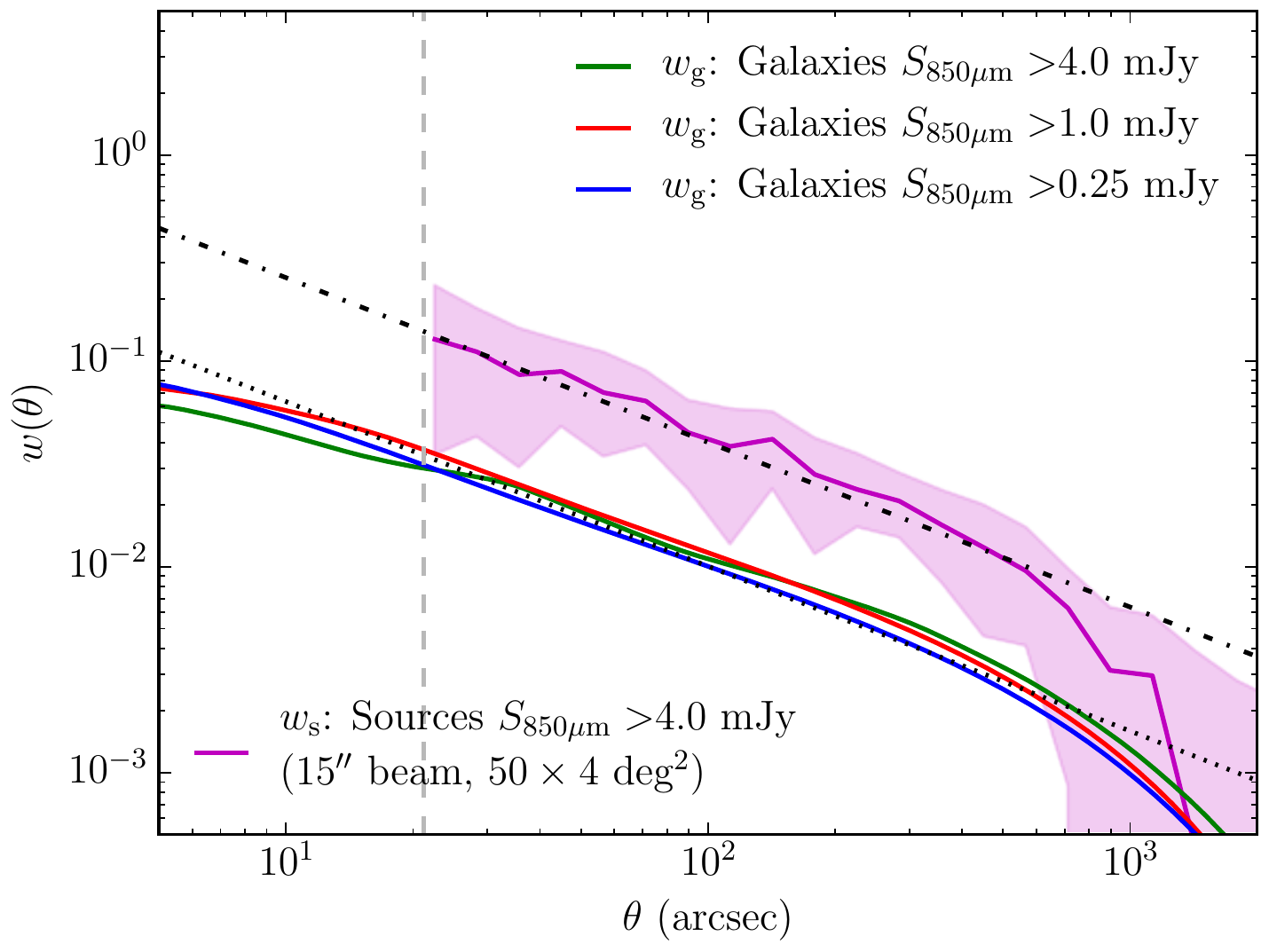}
\caption{The predicted angular correlation function for $850$~$\mu$m selected galaxies ($w_{\rm g}$) with $S_{850 \mu \rm m}>0.25$, $1$ and $4$~mJy (blue, red and green lines respectively).  Also shown is the angular correlation function for sources with $S_{850 \mu \rm m}>4$~mJy extracted from simulated single-dish sub-mm imaging produced with a $15$~arcsec FWHM Gaussian beam (magenta line) with the corresponding shaded region indicating the $1\sigma$ ($16-84^{\mathrm{th}}$ percentile) field-to-field variation over $50$ lightcone realisations of $4$~deg$^{2}$ each.  For reference, the diagonal dotted and dash-dotted lines show two $w\propto\theta^{1-\gamma}$ power laws, with $\gamma=1.8$, offset from each other in amplitude by a factor of $4$.}
\label{fig:wtheta_wg_ws}
\end{figure}

Angular clustering, $w(\theta)$, can be thought of as the on-sky projection of $\xi(r,z)$, weighted by the number density of selected objects at a given redshift.  We therefore use the approximation of \cite{Limber53} to calculate $w_{\rm g}(\theta)$ from $\xi(r,z)$, the spatial two-point correlation function.  This assumes that the selection function (redshift distribution) of galaxies changes slowly over the comoving separations $r$ for which $\xi(r,z)$ is appreciably non-zero. Assuming a flat cosmology (as we do throughout), this allows $w_{\rm g}(\theta)$ to be related to $\xi(r,z)$ by 
\begin{equation}
w_{\rm g}(\theta) = \frac{\int N(z)^{2}\frac{{\rm d}z}{{\rm d}\chi}{\rm d}z\int{\rm d}u\,\xi(r,z)}{[\int N(z){\rm d}z]^{2}}{\rm ,}
\label{eq:limber_wg}
\end{equation}
where $N(z)$ is the predicted redshift distribution of the selected galaxies, ${\rm d}z/{\rm d}\chi = H_{0}E(z)/c$ with $E(z)=[\Omega_{\rm m}(1+z)^3 + \Omega_{\Lambda}]^{1/2}$, $\chi$ corresponds to the comoving radial distance to redshift $z$.  The comoving line of sight separation $u$ is defined by $r=[u^2+\chi^2\varpi^2]^{1/2}$ where $\varpi^2/2 =[1-\cos(\theta)]$.  We present $w_{\rm g}$ for our sub-mm selected galaxy populations, as defined in the previous section, in Fig.~\ref{fig:wtheta_wg_ws}.

\subsection{The Angular Clustering of Single-Dish Sources}
To make predictions for the angular clustering from sub-mm sources that would be observed in single-dish surveys we simulate such observations using the method presented in \cite{Cowley15}.  

Briefly, we generate lightcone catalogues of simulated SMGs using the method described in \cite{Merson13}\footnote{ This does not include any treatment of gravitational lensing.}.  We include in our lightcone catalogue galaxies brighter than the flux at which $90\%$ of the predicted CIB at $850$~$\mu$m is recovered.  The predicted value of the CIB is in good agreement with the observations of \cite{Fixsen98},  and thus gives our image a realistic background.  The galaxies are then binned into pixels according to their on-sky position, with the flux value of a pixel being the sum of the fluxes of all the galaxies within it.  The pixel scale is chosen such that the beam is well sampled.  This image is then smoothed with a Gaussian with a FWHM chosen to be equal to that of the beam used in observational studies following which Gaussian white noise is added of a magnitude comparable to that found in observations.  The image is constrained to have a mean of zero by the subtraction of a uniform background, and then matched-filtered prior to source extraction.  Sources are found by iteratively identifying the maximal pixel in the map and subtracting off the matched-filtered PSF scaled to and centred on the value and position of the pixel.  For simplicity the position of the source is recorded as being at the centre of the identifying pixel.  The result of this source extraction is referred to hereafter as our source-extracted catalogue.

Here we choose to make predictions for the $850$~$\mu$m \mbox{SCUBA-2} Cosmology Legacy Survey \citep[S2CLS, e.g.][]{Geach13}, as measuring the clustering of SMGs is one of the main survey goals.  For this reason we choose a Gaussian beam with a FWHM of $15$~arcsec (similar to that of the SCUBA-2/JCMT configuration at $850$~$\mu$m).  In order to estimate field-to-field variation we generate $50\times4$~deg$^2$ randomly orientated lightcones.  We add instrumental Gaussian white noise with $\sigma_{\rm inst}=1$~mJy/beam, which gives our maps a total noise of $\sigma_{\rm tot}\approx1.2$~mJy/beam, calculated from a pixel histogram of our image.  This broadening of the noise distribution is due to the confusion noise from faint unresolved galaxies in the image, as $\sigma_{\rm tot}\approx\sqrt{\sigma_{\rm inst}^{2}+\sigma_{\rm conf}^{2}}$. We extract sources down to $4$~mJy ($\sim3.5\sigma$) as this is the typical limit at which sources are extracted out of single-dish surveys \citep[e.g.][]{Coppin06,Weiss09}.
  
To calculate $w_{\rm s}$ for our source extracted catalogue we use the angular equivalent of equation (\ref{eq:LandySzalay}).  To ensure we are not affected by noise in the random catalogue, we generate random catalogues using the same selection function as for the data (i.e. same survey geometry) but with $250$ times the number of points as there are sources for each of our simulated surveys.

In estimating $w_{\rm s}(\theta)$ for each of the 50 lightcone realisations we used the actual number of sources in each field to calculate the mean surface density in order to match what is done observationally, rather than the true mean surface density. This causes the mean angular correlation function to be underestimated by an average amount
\begin{equation}
\sigma^2 = \frac{1}{\Omega^{2}}\int\int w_{\rm true}(\theta)\,{\rm d}\Omega_{1}{\rm d}\Omega_{2}{\rm ,}
\end{equation} 
\citep{GrothPeebles77} due to the integral constraint (that by construction the estimated angular correlation function will integrate to zero over the area of the field), where $w_{\rm true}(\theta)$ is the true angular correlation function of the sources and the angular integrations are over a field of area $\Omega$.  This quantity is related to the field-to-field variation in the number counts through
\begin{equation}
\sigma^2 = \frac{\langle(\eta - \langle\eta\rangle)^{2}\rangle}{\langle\eta\rangle^{2}}-\frac{1}{\langle\eta\rangle}{\rm ,}
\label{eq:sigma2}
\end{equation}
\citep[e.g.][]{Efstathiou91}.  We evaluate equation (\ref{eq:sigma2}) for our $50\times 4$~deg${^2}$ lightcones and find $\sigma^2=4.8\times10^{-5}$, which we add onto our computed angular correlation functions for sub-mm sources ($w_{s}$).

\begin{figure}
\includegraphics[width=\columnwidth]{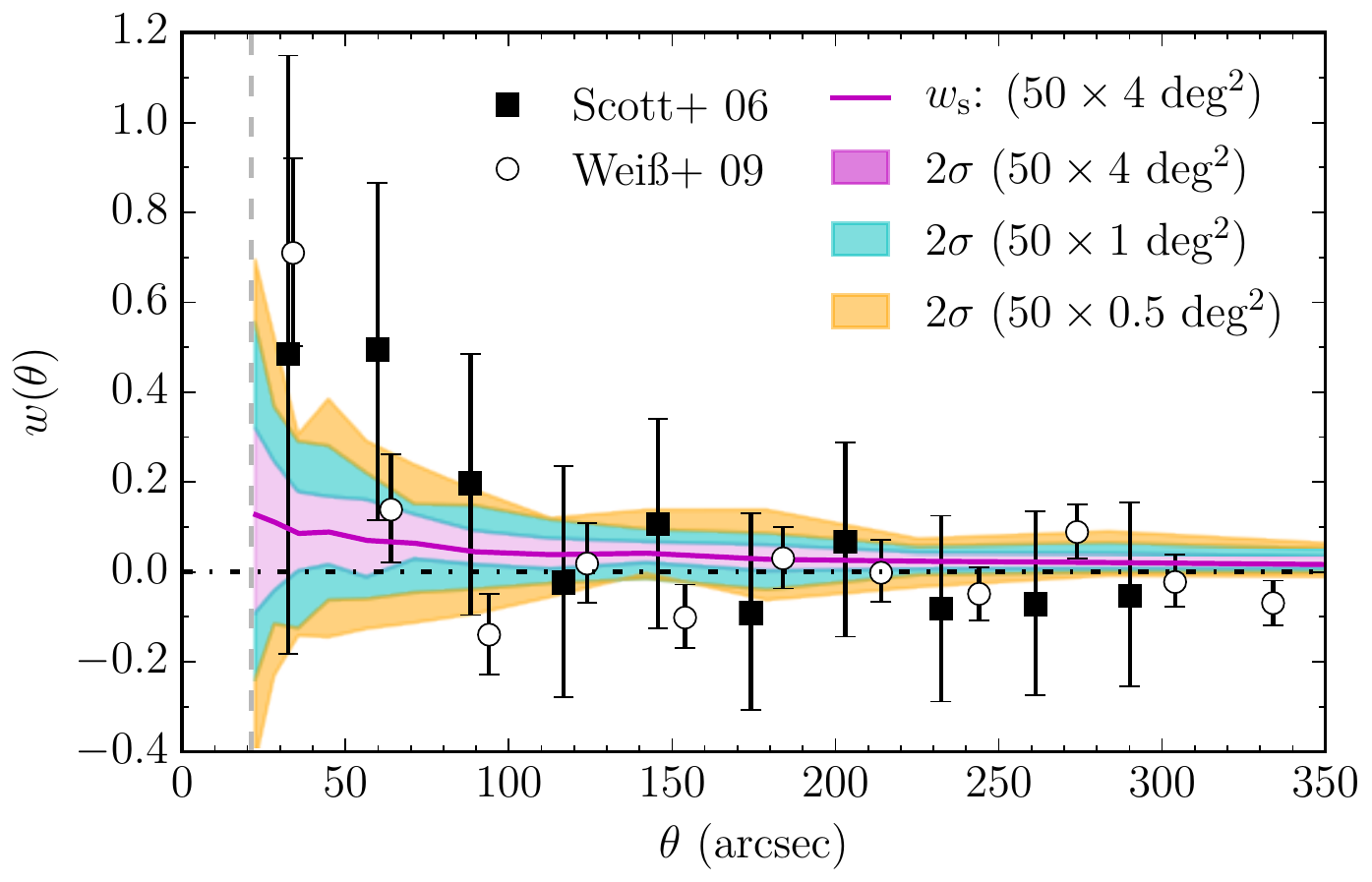}
\caption{Comparison of the predicted angular correlation function for our $S_{850\mu\rm m}>4$~mJy simulated single-dish source catalogue, $w_{\rm s}$ (magenta line), to observational estimates from Scott et al. (\citeyear{Scott06}, filled squares) and Wei{\ss} et al. (\citeyear{Weiss09}, open circles). The shaded magenta, cyan and orange regions indicate the $2\sigma$ ($2.25-97.75^{\rm th}$ percentile) range derived from the field-to-field variation over 50 lightcone realisations for fields of $4$, $1$ and $0.5$~deg$^2$ respectively.}
\label{fig:obs_wtheta}
\end{figure}

In Fig. \ref{fig:wtheta_wg_ws} we show the mean $w_{s}(\theta)$ from the $50$ lightcone realisations (magenta line), with the corresponding shaded region indicating the $1\sigma$ ($16-84^{\rm th}$ percentile) field-to-field variation in $w_{s}(\theta)$ in each bin of angular separation.  In Fig. \ref{fig:obs_wtheta}
 we compare $w_{\rm s}(\theta)$ with observational estimates from the $0.35$~deg$^2$ LESS source catalogue (Wei{\ss} et al. \citeyear{Weiss09}, $19$~arcsec FWHM, $S_{850\mu\rm m}\gtrsim$4.5~mJy); and from sources identified from a compilation of non-contiguous SCUBA fields totalling $\sim0.13$~deg$^2$ in area (Scott et al. \citeyear{Scott06}, $15$~arcsec FWHM, $S_{850\mu\rm m}\gtrsim5$~mJy).  The magenta, cyan and orange shaded regions indicate the $2\sigma$ ($2.25-97.75^{\rm th}$ percentile) field-to-field variation in each bin of angular separation we predict for fields of $4$, $1$ and $0.5$~deg$^{2}$ respectively, which must be considered when comparing theory and observations.  For this we recalculate the angular correlation function for each field considering only sources within the central $1$ or $0.5$~deg$^2$.  As in Fig. \ref{fig:r0_evolution}, the large error bars of the observational data make a detailed comparison difficult and highlight the need for larger sub-mm surveys.  We note however, that our predictions are consistent with the data once field-to-field variations are taken into account.
\subsection{Blending Bias in the Angular Clustering of Single-Dish Sources}
\begin{figure}
\includegraphics[width=0.95\linewidth]{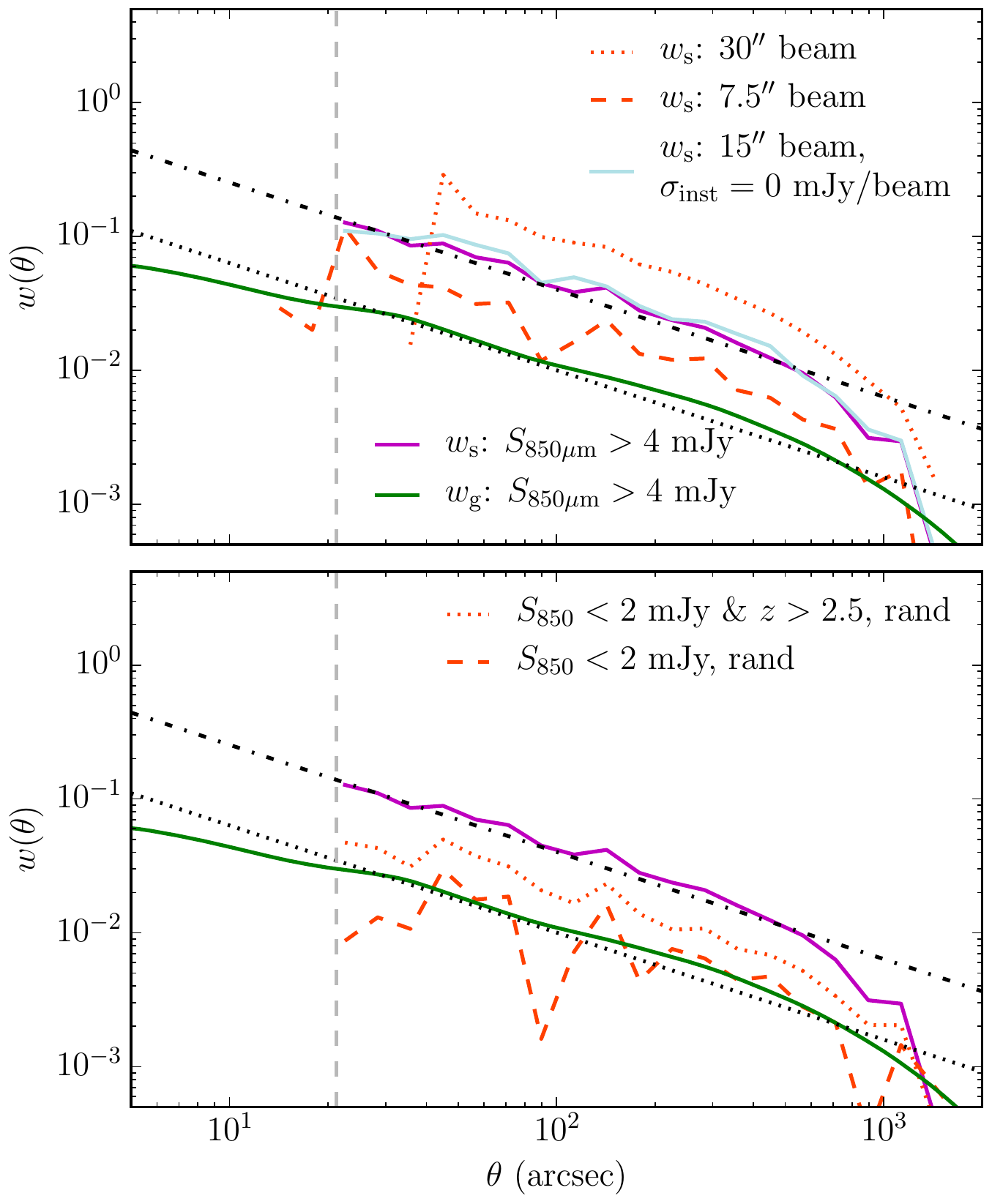}
\caption{The effect of beam-size, instrumental noise and the clustering of faint ($S_{850\mu\rm m}<2$~mJy) galaxies on the angular correlation function of brighter ($S_{850\mu\rm m}>4$~mJy) single-dish sources.  The green and magenta lines show the angular correlation function for galaxies and sources (for a $15$ arcsec beam) respectively, as shown in Fig.~\ref{fig:wtheta_wg_ws}.  The vertical dashed, and diagonal dashed and dash-dotted lines, shown for reference, are also as described in Fig.~\ref{fig:wtheta_wg_ws}.  \emph{Upper panel:}  The dotted (dashed) orange line indicates the correlation function for sources extracted from simulated images generated with a $30$ ($7.5$) arcsec beam.  The light blue line is the source correlation function derived from images created with no `instrumental' noise added. \emph{Lower panel:} The dotted orange line indicates the correlation function for sources extracted from images where the position of galaxies with $S_{850\mu\rm m}<2$~mJy and $z>2.5$ were randomised prior to creation.  The orange dashed line shows the same for images which had the position of all galaxies with $S_{850\mu\rm m}<2$~mJy randomised.}
\label{fig:blending_bias_tests}
\end{figure}
\begin{figure}
\includegraphics[width=0.95\linewidth]{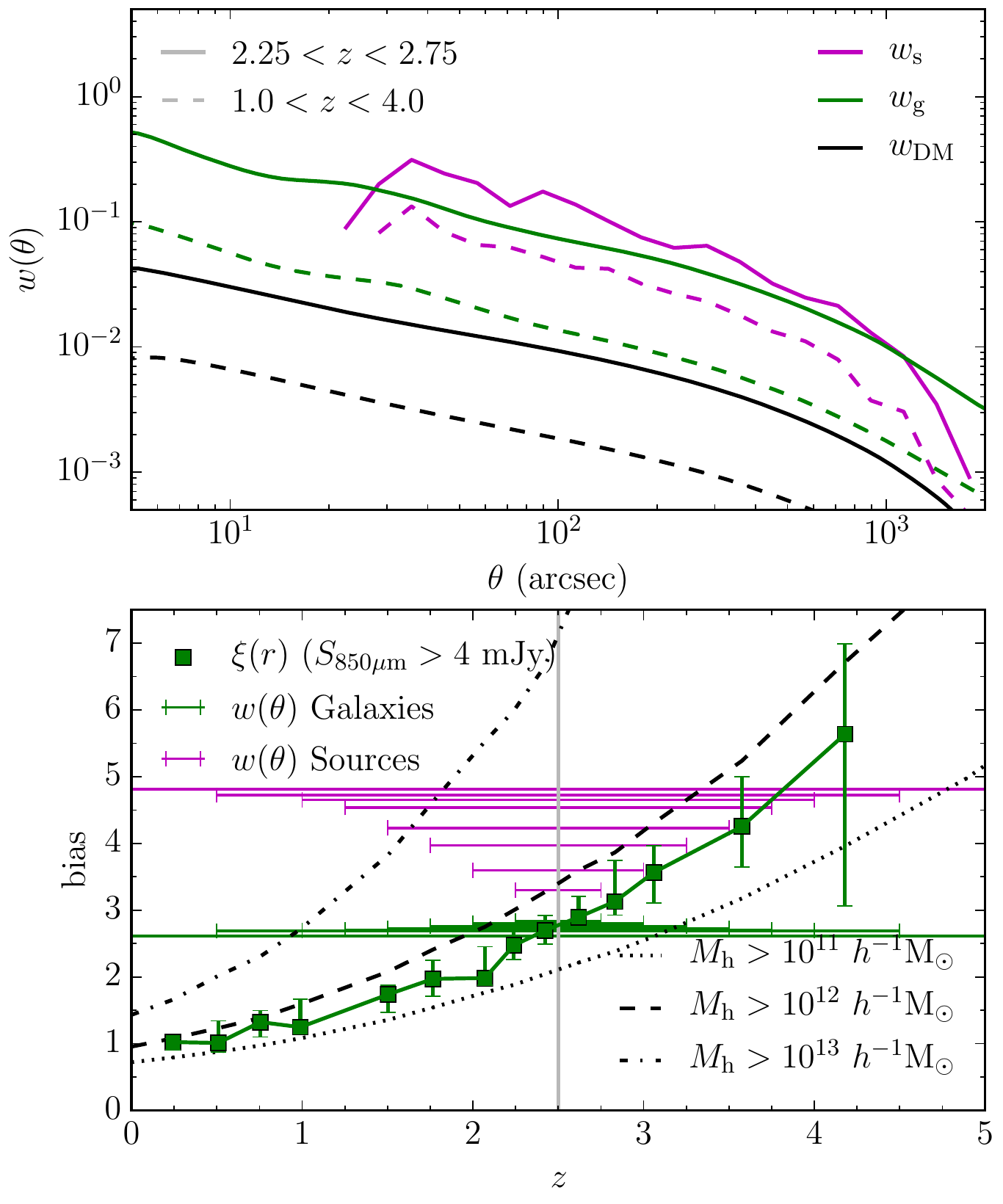}
\caption{The effect of the redshift interval considered on the angular correlation function of $S_{850\mu\rm m}>4$~mJy single-dish source counterparts (see text).  \emph{Upper panel:}  Angular correlation function of single-dish source counterparts (magenta lines), $S_{850\mu\rm m}>4$~mJy galaxies (green lines) and dark matter (black lines) for the redshift interval $2.25<z<2.75$ (solid lines) and $1.0<z<4.0$ (dashed lines).  \emph{Bottom panel:}  Evolution of large scale bias with redshift.  Green squares indicate the bias evolution of $S_{850\mu\rm m}>4$~mJy galaxies, derived from the spatial correlation function as in Fig. \ref{fig:r0_evolution}.  The dotted, dashed and dash-dotted lines indicate the bias evolution of halos with $M_{\mathrm{halo}}>10^{11}$, $10^{12}$ and $10^{13}$~$h^{-1}$~M$_{\odot}$ respectively.  The green bars indicate the bias derived from the angular correlation functions of galaxies and dark matter, with the width of the bar indicating the redshift interval considered.  The magenta bars show the same but for bias derived from the angular correlation functions of single-dish source counterparts.  The vertical grey line indicates $z=2.5$, on which all redshift intervals considered are centred.}
\label{fig:blending_bias_tests_z}
\end{figure}
\begin{figure}
\includegraphics[width=0.95\linewidth]{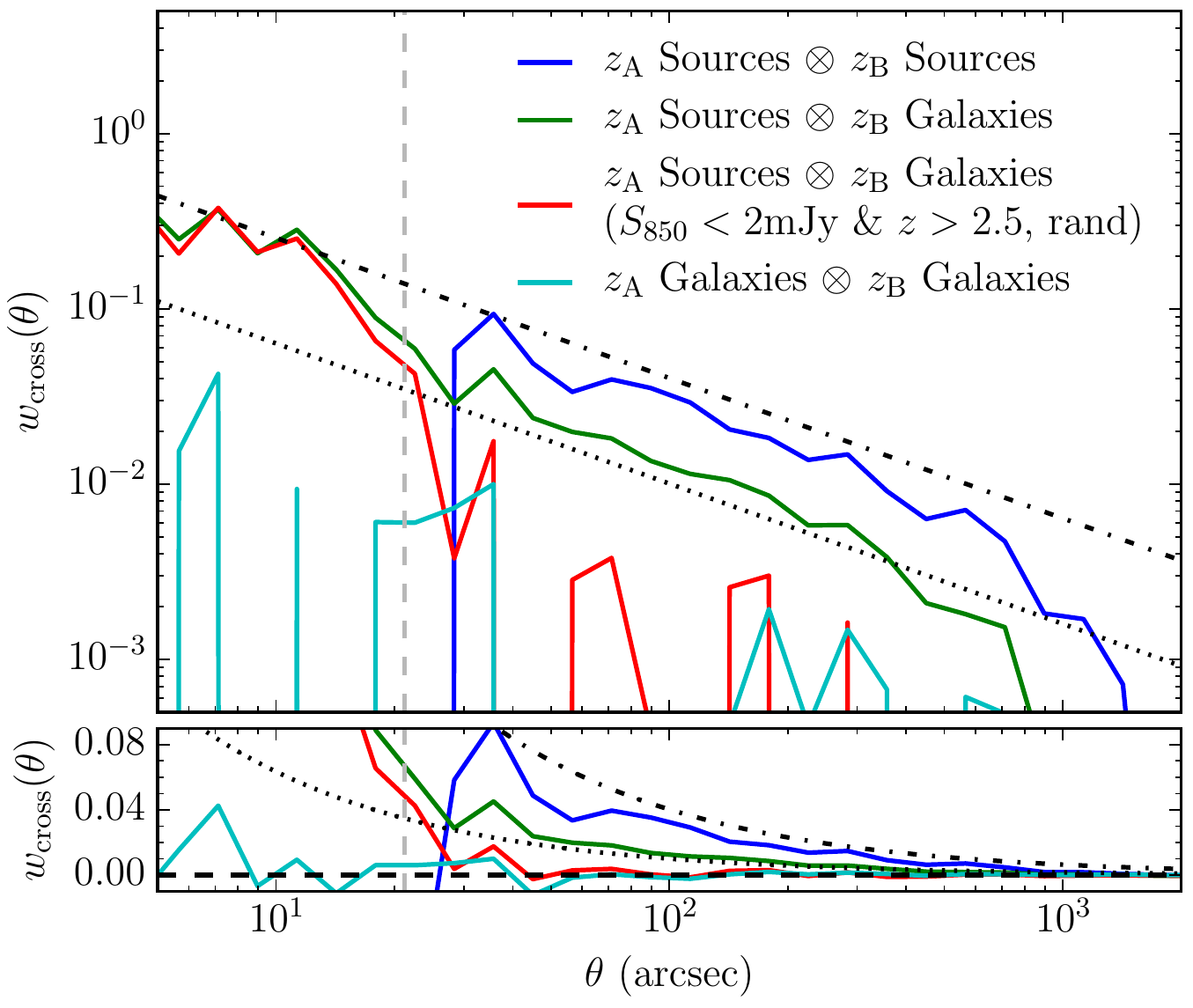}
\caption{Angular cross correlations between two separated redshift intervals, $z_{\rm A}=[1.0,2.4)$ and $z_{\rm B}=[2.6,4.0)$.  In the legend `Sources' refers to the counterparts of sources (see text) extracted from our simulated imaging with $S_{850\mu\rm m}>4$~mJy and `Galaxies' refers to galaxies selected with $S_{850\mu\rm m}>2$~mJy.  \emph{Top panel}:  We show the angular cross correlation of: (i) Source counterparts in $z_{\rm A}$ with source counterparts in $z_{\rm B}$ (blue line); (ii) Source counterparts in $z_{\rm A}$ with galaxies ($S_{850\mu\rm m}>2$~mJy) in $z_{\rm B}$ (green line); (iii) Source counterparts in $z_{\rm A}$ with galaxies ($S_{850\mu\rm m}>2$~mJy) in $z_{\rm B}$ but with the sources extracted from images where the positions of galaxies with $S_{850\mu\rm m}<2$~mJy and $z>2.5$ were randomised prior to creating the images (red line); and (iv) Galaxies ($S_{850\mu\rm m}>2$~mJy) in $z_{\rm A}$ with galaxies ($S_{850\mu\rm m}>2$~mJy) in $z_{\rm B}$ (cyan line).  The vertical dashed, and diagonal dashed and dash-dotted lines, shown for reference, are as described in Fig. \ref{fig:wtheta_wg_ws}. \emph{Bottom panel}: As for top panel but with a linear $y$-axis.  A dashed line at $w=0$ has been added for reference.}
\label{fig:blending_bias_xcorr}
\end{figure}
One of the key results of this work, evident in Fig.~\ref{fig:wtheta_wg_ws}, is that the angular correlation function of sources, $w_{\rm s}$, is greater in amplitude by a factor of $\sim4$ than the angular correlation function of galaxies, $w_{\rm g}$, for the source flux limit used here ($4$~mJy).  In this section we investigate the dependence of this effect on a number of factors and conclude that is to due to confusion in the simulated survey caused by the $15$~arcsec FWHM beam blending the emission of multiple, typically physically unassociated galaxies \citep{Cowley15}, with an on-sky separation comparable to or less than the size of the beam into an object recognised as a single source by the source extraction algorithm\footnote{In \cite{Cowley15} we showed that this confusion effect boosts the cumulative $850$~$\mu$m number counts by a factor of $\sim2$ at $S_{850\mu\rm m}=4$~mJy for a $15$~arcsec FWHM beam.  See also \cite{Hayward13b} and \cite{Munoz15} who investigate the effect of coarse angular resolution on the observed sub-mm number counts.}.  Thus the angular distribution of sources found in the simulated map is different from the angular distribution of the input galaxies.  We label this effect `blending bias,' $b_{\rm b}$, where $b_{\rm b}^2\equiv[w_{s}(\theta)/w_{g}(\theta)]$, and note that a similar effect has been observed in low resolution X-ray surveys \citep[e.g.][]{VikhlininForman95,Basilakos05}.

In the upper panel of Fig.~\ref{fig:blending_bias_tests} we test how sensitive this bias is to the size of the beam and `instrumental' noise.  We repeat the calculation for deriving the angular correlation function of single-dish sources for images generated using Gaussian beams with FWHM of $30$ and $7.5$~arcsec.  We kept the instrumental noise constant at $\sigma_{\rm inst}=1$~mJy/beam in each case and used the same flux limit of $S_{850\mu\rm m}>4$~mJy to select our sources, noting that varying the beam size will change the confusion in the image and thus the overall noise.  We derived blending bias factors in $w_{\rm s}$ of $b_{\rm b}^2\sim2$ and $b_{\rm b}^2\sim8$ for the $7.5$ and $30$~arcsec beams respectively.  We tested the effect of instrumental noise by creating a set of images with a $15$~arcsec beam, but without the addition of instrumental noise.  This can be seen in Fig.~\ref{fig:blending_bias_tests} to have a negligible effect on the angular correlation function of the sources, as one would expect given that our `instrumental' noise is random and has no dependence on scale.  

In the lower panel of Fig.~\ref{fig:blending_bias_tests} we repeat the calculation on images which had the positions of galaxies with $S_{850\mu\rm m}<2$~mJy and $z>2.5$ randomised prior to being created and find that the blending bias is reduced to $b_{\rm b}^2\sim2$.  For maps which had the position of all galaxies with $S_{850\mu\rm m}<2$~mJy randomised the blending bias is approximately unity i.e. has been removed.  Although not shown in Fig.~\ref{fig:blending_bias_tests}, we also tested this effect on a set of images which had the positions of all galaxies randomised prior to their creation, and observed a result consistent with the selected sources being completely unclustered.  We conclude that blending bias in the angular clustering of single-dish sources is due to the confusion noise or rather the clustering of faint unresolved galaxies and the way in which, when their emission is smoothed with a single-dish beam, this causes certain on-sky positions to be selected as sources.  It thus depends on the combined effect of the finite beam size, the intrinsic clustering of the underlying galaxies, and their intrinsic number counts.

We also consider how calculating the angular correlation function using different redshift intervals can affect the blending bias.  In order to assign a redshift to a single-dish source we first define a source-counterpart as the galaxy which is contributing the most sub-mm flux to a source, taking into account the profile of the beam.  We can then select these counterparts within a given redshift interval and recalculate the angular correlation function, now using the on-sky position of the counterpart.  For the underlying galaxies and dark matter we calculate the angular correlation function over a given redshift interval by appropriately changing the limits in the \cite{Limber53} equation (\ref{eq:limber_wg}).  An example of this is shown in the upper panel of Fig.~\ref{fig:blending_bias_tests_z} for two redshift intervals centred on $z=2.5$, $2.25<z<2.75$ (solid lines) and $1.0<z<4.0$ (dashed lines).  In this way we can derive a large-scale bias, defined as $[w(\theta)/w_{\rm DM}(\theta)]^{1/2}$, for the galaxies and source-counterparts, as a function of redshift interval considered.  This is shown in the bottom panel of Fig.~\ref{fig:blending_bias_tests_z} where we consider $8$ redshift intervals of varying width centred on $z=2.5$.  We can see that the derived source-counterpart bias, which is affected by blending bias, increases monotonically as the width of the redshift interval increases whilst the bias derived from the angular correlation function of galaxies is approximately constant and consistent with the bias derived from the spatial correlation function (see Section \ref{subsec:spatial_SMGs}) for all redshift intervals considered.  Also evident in this panel is how the halo mass can be significantly over-estimated as a result of this effect.  As a further example of this, using equation (8) in \cite{SMT01} to infer halo mass from a measured bias, we find that doubling the bias (i.e. $b_{\rm b}=2$) of halos with mass $10^{12}$~$h^{-1}$~M$_{\odot}$ yields an inferred halo mass of $10^{13.1}$~$h^{-1}$~M$_{\odot}$ at a redshift of $2.5$, an over-estimation of more than an order of magnitude.

To further illustrate the results in this section we imagine a simplified scenario with two distinct redshift intervals A and B and two angular positions $\underline{\theta}_{1}$ and $\underline{\theta}_{2}$.  Within each redshift interval the positions of galaxies will be correlated according to some $w(S_{1},S_{2},z\pm\Delta z,|\underline{\theta}_{1}-\underline{\theta}_{2}|)$, and we define some flux limit $S_{\mathrm{lim}}$ brighter than which galaxies will be resolved as point sources in the beam-smoothed imaging and fainter than which they would require some boost to be counted in the single-dish catalogue.

If we now consider the effect of the beam, we have a beam-smoothed flux density field in each redshift interval, $\mathcal{S}(\Omega_{\rm beam},z\pm\Delta z,\underline{\theta})$, dominated by galaxies with $S<S_{\rm lim}$, the distribution of which will be correlated with the positions of galaxies with $S>S_{\rm lim}$ in that interval, according to $w$.  It is also now possible for flux from B to boost objects (at the same on-sky position) in A into the selection (and vice-versa).  This induces an artificial cross-correlation between the sources selected in A and B, as some objects in B required a flux boost from A to be considered and this flux is correlated with selected objects in A.  Thus we make the prediction that the cross-correlation of single-dish source counterparts (for sources with $S_{850\mu\rm m}>4$~mJy) in distinct redshift intervals will be non-zero, even in the absence of effects such as gravitational lensing which are not considered here.  

This is demonstrated in Fig.~\ref{fig:blending_bias_xcorr}, where we show the angular cross correlation between source counterparts in two distinct redshift intervals  $1.0\leq z<2.4$, $z_{\rm A}$, and $2.6\leq z<4.0$, $z_{\rm B}$ (blue line).  This is found to be non-zero whilst the equivalent calculation for bright galaxies (with $S_{850\mu\rm m}>2$~mJy\footnote{Here we use a limit of $2$, rather than $4$ mJy, so we have enough objects for a robust determination of $w_{\rm cross}$.  We do not expect the result to be sensitive to this given that the auto-correlation of galaxies is roughly independent of flux over this flux range.}) is zero (cyan line).  We also find that source counterparts in $z_{\rm A}$ are correlated with bright galaxies in $z_{\rm B}$, in this case shown for galaxies with $S_{850\mu\rm m}>2$~mJy  (green line).  The physical correlation of the faint with the bright galaxies in $z_{\rm B}$ has caused the sources from $z_{\rm A}$, many of which were selected as sources because of a flux contribution from faint galaxies in $z_{\rm B}$, to be correlated with bright galaxies in $z_{\rm B}$.  This is an induced correlation introduced by the finite beam.  When we repeat the source-galaxy cross-correlation using sources from maps which had the positions of galaxies with $S_{850\mu\rm m}<2$~mJy and $z>2.5$ randomised prior to the image being created, the randomisation removes the physical correlation between faint and bright galaxies in $z_{\rm B}$, thus we find that the induced cross-correlation between sources in $z_{\rm A}$ and bright galaxies in $z_{\rm B}$, on scales larger than the beam, is now zero.  This is despite the fact the positions of galaxies with $S_{850\mu\rm m}>2$~mJy in $z_{\rm B}$ were not changed. 

We infer that it is these induced cross-correlations that cause the trend in blending bias with redshift interval width seen in the lower panel of Fig.~\ref{fig:blending_bias_tests_z}, as increasing the redshift interval increases the number of induced cross-correlations considered.  It also explains the trends seen in the lower panel of Fig.~\ref{fig:blending_bias_tests}, as randomising the positions of faint galaxies reduces the correlation between the distribution of flux density, $\mathcal{S}$, and the distribution of galaxies with $S>S_{\rm lim}$ at a given redshift, and thus the contribution of the induced cross-correlation terms.  For the same $S_{\rm lim}$ increasing the beam-size will on average increase the multiplicity of sources.  As the components of each source are, in our simulations, drawn from different redshift intervals (galaxies composing a single source are generally at different redshifts) this means that for each source more induced cross-correlation terms are considered, producing the trends seen in the upper panel of Fig.~\ref{fig:blending_bias_tests}.    

We therefore caution that significant modelling is needed to interpret the angular correlation function of sources identified in single-dish surveys, at flux limits at which the sources are confused (i.e. composed of multiple fainter galaxies).  The implication is that the halo masses of the galaxies in question could be seriously overestimated if blending bias is not corrected for.  It appears from Fig.~\ref{fig:wtheta_wg_wI} that $w_{\rm I}(\theta)$, described in the next section, exhibits angular clustering more representative of the underlying galaxy population. We suggest then that information regarding the halo masses of SMGs should be inferred from $w_{\rm I}(\theta)$.  This comes with the important caveat that the effects of correlated noise in observed images, e.g. large-scale structure due to correlated atmospheric contamination and $1/f$ noise, need to be removed or accurately modelled.

Targeted follow-up of single-dish sources with interferometers could also be used to overcome blending bias, as the order of magnitude better resolution would allow the underlying galaxies from which the sources are composed to be identified, down to flux limits dependent on integration time.  This would provide an approximately complete flux-limited catalogue of galaxies down to slightly above the source-extraction limit of the single-dish survey \citep[some galaxies are de-boosted by instrumental noise to below the flux limit of the single dish survey and are therefore missed from the follow-up observations, e.g.][]{Karim13,Cowley15} which could then be used to derive the correlation function free from blending bias. 
 
\subsection{The Angular Clustering of Intensity Fluctuations}
\begin{figure}
\includegraphics[width=\columnwidth]{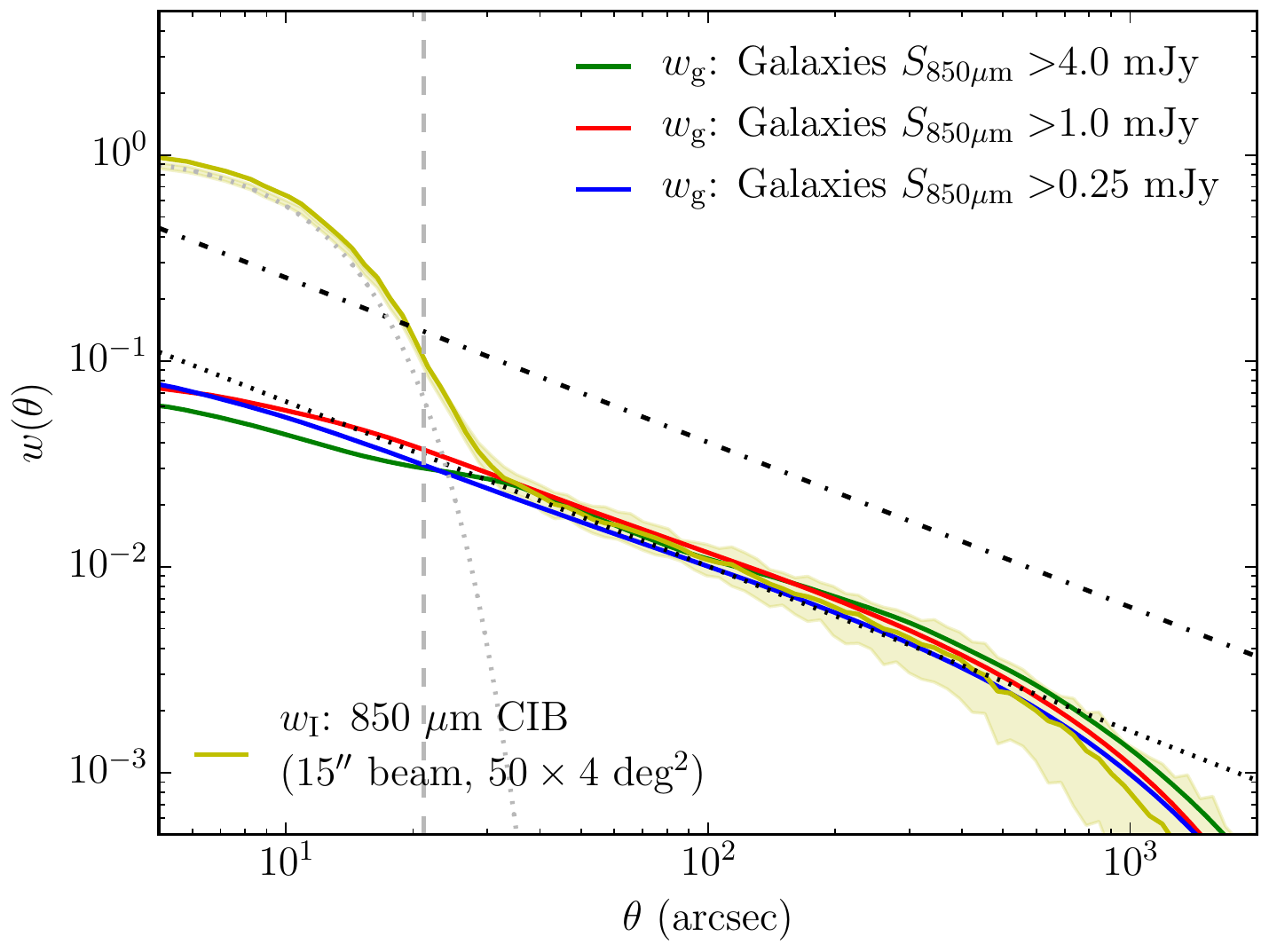}
\caption{Predicted angular auto-correlation functions.  The angular correlation function of the $850$~$\mu$m intensity fluctuations, derived from the angular power spectrum of the simulated single-dish imaging, prior to matched filtering is shown by the gold line. The gold shaded region indicates the $1\sigma$ ($16-84^{\mathrm{th}}$ percentile) field-to-field variation over $50$ lightcone realisations of $4$~deg$^{2}$ each. The grey dotted line indicates the expectation for the angular correlation function of the CIB intensity fluctuations if the galaxies contributing to it were unclustered.  All other lines are as described in Fig.~\ref{fig:wtheta_wg_ws}}
\label{fig:wtheta_wg_wI}
\end{figure}
In this section\footnote{In this section, for ease of reading, and as here we are only considering a single band ($850$~$\mu$m), we suppress the explicit frequency dependence in our notation.  For example, we write the mean intensity at a given observed frequency $\nu$,  $\langle I_\nu\rangle$, as $\langle I\rangle$.} we calculate the angular clustering of intensity fluctuations in our simulated images, $w_{\rm I}(\theta)$.  We first introduce this quantity before describing how it is calculated in this paper.  It can be defined as
\begin{equation}
\langle I(\underline{\theta}_{1})I(\underline{\theta}_{2})\rangle = \langle I\rangle^{2}[1+w_{\rm I}(\theta)]{\rm ,}
\end{equation} where $I(\underline{\theta}_{1})$ represents the intensity in a given direction $\underline{\theta}_{1}$,  $\theta = |\underline{\theta}_{1}-\underline{\theta}_{2}|$ and $\langle I\rangle$ is the mean intensity, which can be calculated from the number counts of our model by
\begin{equation}
\langle I\rangle = \int S \frac{\mathrm{d}\eta}{\mathrm{d}S} \mathrm{d}S\mathrm{.}
\label{eq:Inu}
\end{equation}  
The function $w_{\rm I}(\theta)$ can be expressed as a flux-weighted integral of the angular correlation function of galaxies, $w_{\rm g}$, such that
\begin{equation}
\begin{split}
w_{\rm I}(\theta) &=& \frac{1}{\langle I\rangle^{2}}\left[\int\int w_{\rm g}(S_{1},S_{2},\theta)S_{1}S_{2}\frac{{\rm d}\eta}{{\rm d}S_{1}}\frac{{\rm d}\eta}{{\rm d}S_{2}}{\rm d}S_{1}{\rm d}S_{2}\right.\\
& & +\left.\delta^{\rm D}(\theta)\int S^{2}\frac{{\rm d}\eta}{{\rm d}S}{\rm d}S\right]
\end{split}
\label{eq:wI_int}
\end{equation}
where $w_{\rm g}(S_{1},S_{2},\theta)$ is the angular cross-correlation of galaxies with fluxes $S_{1}$ and $S_{2}$ and ${\rm d}\eta/{\rm d}S_{i}$ is the surface density per unit solid angle of galaxies with flux $S_{i}$.  The angular cross-correlation of galaxies $w_{\rm g}(S_{1},S_{2},\theta)$ derives from a more general form of equation (\ref{eq:limber_wg}) such that
\begin{equation}
w_{\rm g}(S_{1},S_{2},\theta) = \frac{\int N_{1}(z)N_{2}(z)\frac{{\rm d}z}{{\rm d}\chi}{\rm d}z\int{\rm d}u\,\xi(S_{1},S_{2},r,z)}{\int N_{1}(z){\rm d}z\int N_{2}(z){\rm d}z},
\end{equation}
where $N_{i}(z)$ represents the redshift distribution of galaxies with flux $S_{i}$ and $\xi(S_{1},S_{2},r,z)$ is the spatial cross-correlation of galaxies with $S_{1}$ and $S_{2}$, at redshift $z$.  We can recover $w_{\rm g}$ for an individual galaxy population by integrating $w_{\rm g}(S_{1},S_{2},\theta)$ over the flux limits defining the selection of the population.  The term containing the Dirac delta function $\delta^{\rm D}(\theta)$ on the right hand side of equation (\ref{eq:wI_int}) is the shot noise, which arises from galaxies being approximated as point sources.

We can calculate $w_{\rm I}$ for the clustered galaxy population directly from our simulated images using the estimator 
\begin{equation}
w_{\rm I}(\theta) = \frac{\sum_{ij}\delta_{i}\delta_{j}\Theta_{ij}}{\sum_{ij}\Theta_{ij}}{\rm ,}
\label{eq:pix_ACF_estimator}
\end{equation} where $\delta_{i}$ is the fractional variation of flux in the $i^{\rm th}$ pixel and is  calculated using $\delta_{i} = (S_{i}/\langle{S}\rangle)-1$ where $S_{i}$ is the flux value of the $i^{\rm th}$ pixel and $\langle S\rangle$ is the average flux value of a pixel, as all of our pixels are of equal area.  The step function $\Theta_{ij}$ is $1$ if pixels $i$ and $j$ are separated by a distance in the angular bin $\theta\pm\Delta\theta/2$ and zero otherwise.  However, in practice it is more computationally efficient to make use of the fact that $w_{\rm I}$ can be obtained from the angular power spectrum of CIB anisotropies, $P_{\rm I}(k_{\theta})$, using a Fourier transform such that
\begin{equation}
w_{I}(\theta) = \frac{2\pi}{\langle I\rangle^{2}}
\int P_{\rm I}(k_{\theta})J_{0}(2\pi k_{\theta}\theta)k_{\theta}{\rm d}k_{\theta}{\rm ,}
\label{eq:wI_FT}
\end{equation} 
where $J_{0}$ is the zeroth order Bessel function of the first kind and the convention $k_{\theta}=1/\lambda_{\theta}$ is used\footnote{We use this convention as it is the standard practice for angular power spectra of CIB anisotropies \citep[e.g.][]{Gautier92,Viero09}.  Under this convention the angular wavenumber is related to the multipole index, $\ell$, by $\ell=2\pi k_{\theta}$ (when angles are measured in radians).}.  We therefore compute $P_{\rm I}(k_{\theta})$ directly from our simulated images, prior to any matched-filtering, and make use of equation (\ref{eq:wI_FT}) to calculate $w_{\rm I}$.  This quantity is shown in Fig.~\ref{fig:wtheta_wg_wI} (gold line), with the corresponding shaded region indicating the $1\sigma$ percentile variation of our $50$ lightcone realisations at a given $\theta$.  The Gaussian-like profile on small scales ($\theta<30$~arcsec) is due to the beam used to convolve the simulated image and is mostly produced by the shot noise term in equation (\ref{eq:wI_int}).  It can be seen that on scales larger than the beam $w_{\rm I}$ is very similar to $w_{\rm g}$, which is unsurprising given that $\sim70\%$ of the total background light predicted by the model at $850$~$\mu$m is in galaxies with $S_{850\mu\rm m}>0.25$~mJy.

\section{Angular Power Spectrum of CIB Anisotropies}
\begin{figure}
\includegraphics[width=\linewidth]{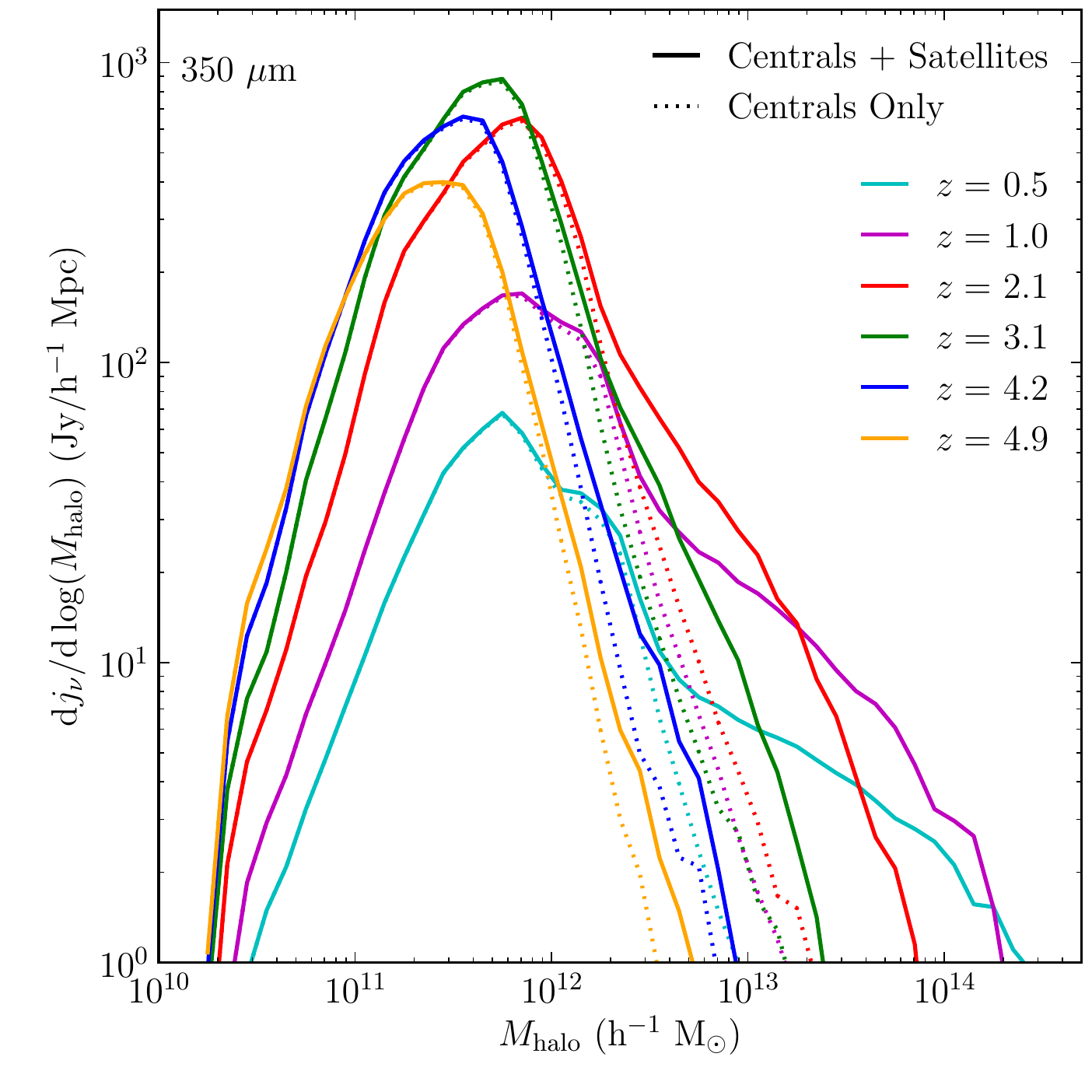}
\caption{Predicted differential emissivity of our model at $350$~$\mu$m for a range of redshifts, as indicated in the legend.  The contribution from central (central + satellite) galaxies is shown using dotted (solid) lines.}
\label{fig:jnu}
\end{figure}
\begin{figure*}
\includegraphics[width=\linewidth]{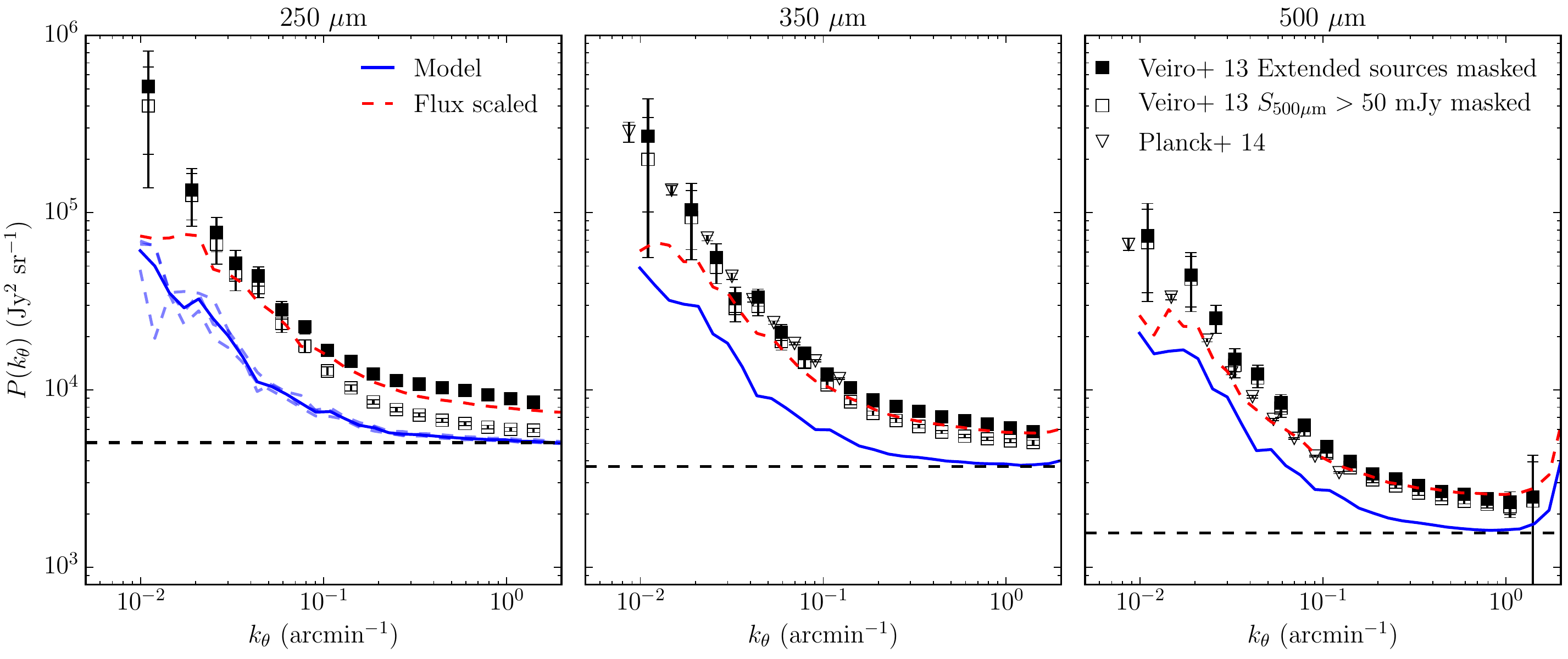}
\caption{Angular power spectra of CIB anisotropies predicted by our model at $250$, $350$ and $500$~$\mu$m (left to right panels).  The blue solid line indicates the power spectrum averaged over $3$ randomly orientated lightcones, each with an area of $20$~deg$^2$. The dashed blue lines in the left panel indicate the power spectra for each of these fields individually.  The horizontal dashed line shows the predicted shot noise contribution to power spectra.  The dashed red line shows the prediction of our model after the fluxes of our simulated galaxies have been rescaled (see text).  We compare to the observational data of Veiro et al. (\citeyear{Viero13}, squares) with the filled and open squares corresponding to different levels of masking, and to that of the Planck Collaboration et al. (\citeyear{Planck14CIB}, triangles).}
\label{fig:SPIRE_Pk}
\end{figure*}
\begin{figure}
\includegraphics[width=\columnwidth]{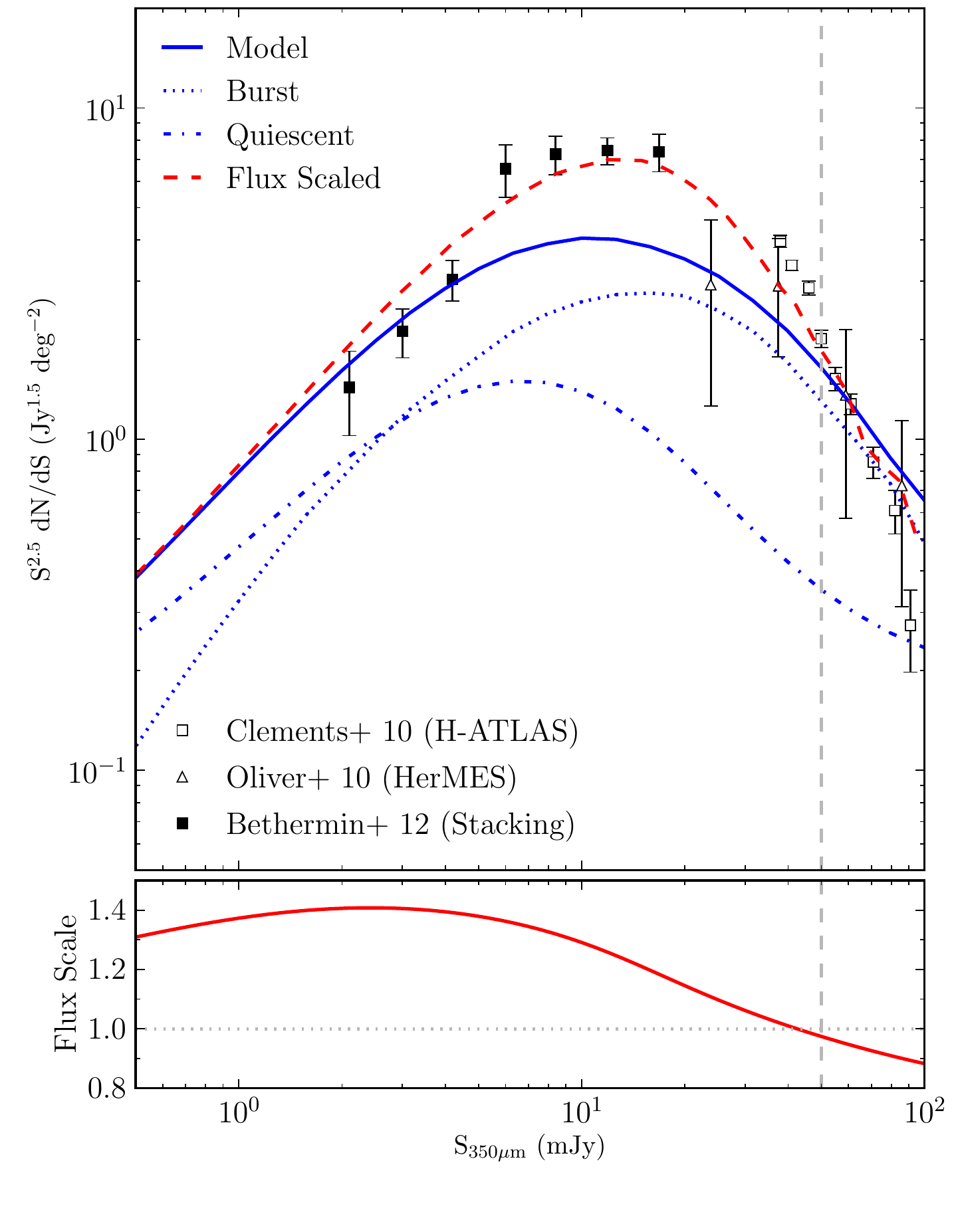}
\caption{An example of our flux re-scaling scheme at $350$~$\mu$m.  \emph{Top panel:} Predicted number counts (blue line) showing the contribution to the counts from starburst and quiescent galaxies (dotted and dot-dashed lines respectively).  The red dashed line shows the number counts after the flux rescaling has been applied.  Observational data are taken from Clements et al. (\citeyear{Clements10}, open squares), Oliver et al. (\citeyear{Oliver10}, open triangles) and Bethermin et al. (\citeyear{Bethermin12}, filled squares).  \emph{Bottom panel:} The flux rescaling applied to simulated galaxies as a function of original model flux.  A horizontal dotted line is drawn at unity for reference.  The vertical dashed line in both panels indicates a flux of $50$~mJy, the limit brighter than which we do not include galaxies in our image in order to match the masking applied by Viero et al. (\citeyear{Viero13}).} 
\label{fig:FluxScale}
\end{figure} 
\label{sec:Pk}
The galaxies which contribute to the bulk of the CIB cannot be individually resolved with current instruments, and instead information regarding their clustering and hence the masses of the halos they occupy is derived from observations of the clustering of fluctuations in the background light.  Therefore, in this section we compare predictions with recent measurements of the angular power spectrum of CIB anisotropies $P^\nu_{\rm I}({k_{\theta})}$. Here $\nu$ is a fixed observed frequency [related to the emitted frequency, $\nu_{\rm e}$, by $\nu=\nu_{\rm e}(1+z)^{-1}$].  

The angular power spectrum of CIB anisotropies was introduced in equation (\ref{eq:wI_FT}) and can be expressed as an integral over redshift of the 3D power spectrum of fractional emissivity fluctuations $\mathcal{P}_{j}^{\nu}(k,z)$, where spatial wavenumber $k$ is related to spatial wavelength $\lambda$ by the convention $k = 2\pi/\lambda$.  Using the approximation of \cite{Limber53}, the small-angle approximation ($k_{\theta}\gg1$) and assuming a flat cosmology, we can write $P^\nu_{\rm I}({k_{\theta})}$ (for $k_{\theta}$ in units of radians$^{-1}$) as
\begin{equation}
P^{\nu}_{\rm I}(k_{\theta}) = \int {\rm d}z \frac{{\rm d}\chi}{{\rm d}z}\left(\frac{a}{\chi}\right)^{2}\langle{j}_{\nu}(z)\rangle^2\mathcal{P}_{j}^{\nu}(k=2\pi k_{\theta}/\chi ,z)
\label{eq:define_Pk}
\end{equation}
\citep[e.g.][]{Viero09,Shang12}. Here $\chi$ is the radial comoving distance to redshift $z$, $a = (1+z)^{-1}$ is the cosmological scale factor and $\langle j_{\nu}(z)\rangle$ describes the mean emissivity per unit solid angle at redshift $z$, which can be expressed as
\begin{equation}
\langle{j}_{\nu}(z)\rangle = 
\int {\rm d}L_{\nu} \frac{{\rm d}n}{{\rm d}L_{\nu}}(L_{\nu},z) \left(\frac{L_{\nu}}{4\pi}\right){\rm ,}
\end{equation}
and related to the mean intensity (see equation \ref{eq:Inu}) by
\begin{equation}
\langle I_{\nu}\rangle = \int {\rm d}z \frac{{\rm d}\chi}{{\rm d}z}a\langle j_{\nu}(z)\rangle{\rm .}
\end{equation}

In our model, not all halos contribute equally to $\langle{j}_{\nu}(z)\rangle$.  We can therefore define a differential emissivity ${\rm d}j_{\nu}/{\rm d}\log_{10} M_{\rm h}$ \citep[e.g.][]{Shang12,Bethermin13} such that equation (\ref{eq:define_Pk}) can be expressed as 
\begin{equation}
\begin{split}
P^{\nu}_{\rm I}(k_{\theta}) = & \int\int\int{\rm d}z\,{\rm d}\log_{10}M_{\rm h}\,{\rm d}\log_{10}{M_{\rm h}}^{\prime}\frac{{\rm d}\chi}{{\rm d}z}\left(\frac{a}{\chi}\right)^{2} \\ 
& \times \frac{{\rm d}j_{\nu}}{{\rm d}\log_{10}M_{\rm h}}\frac{{\rm d}j_{\nu}}{{\rm d}\log_{10}M_{\rm h}^{\prime}}\mathcal{P}_{j}^{\nu}(k,M_{\rm h},M_{\rm h}^{\prime},z){\rm ,}
\end{split}
\label{eq:define_Pk_Mh}
\end{equation}
where $\mathcal{P}_{j}^{\nu}(k,M_{\rm h},M_{\rm h}^{\prime},z)$ is the 3D cross-spectrum of fractional emissivity fluctuations, between halos of mass $M_{\rm h}$ and $M_{\rm h}^{\prime}$.  

Whilst in principle it is possible to calculate $\langle j_{\nu}(z)\rangle$ and $\mathcal{P}^{\nu}_{j}(k,z)$ from the output of our model, for simplicity we compute $P^{\nu}_{\rm I}(k_{\theta})$ from a simulated image of a lightcone catalogue at the wavelength of interest.  

Here, as we compare $P^{\nu}_{\rm I}(k_{\theta})$ predicted by the model to recent \emph{Herschel}-SPIRE data \citep{Viero13}, we use wavelengths of $250$, $350$ and $500$~$\mu$m, and a Gaussian beam with a FWHM of $18$, $25$ and $36$~arcsec respectively, to create our imaging.  For simplicity we do not add any instrumental noise to these maps.  Following the procedure outlined earlier we generate a lightcone catalogue including galaxies brighter than the flux at which we recover 90 percent of the predicted CIB at the wavelength of interest (this predicted CIB agrees well with the observations of \cite{Fixsen98} at all wavelengths) and choose a pixel scale such that the beam is well sampled.  We generate $3\times20$~deg$^{2}$ lightcones in order to have a similar total area to that used by Viero et al.       

First, we show the differential emissivity of our model (described above) at $350$~$\mu$m in Fig.~\ref{fig:jnu}, in terms of the contribution from central and satellite galaxies.  The contribution from central galaxies peaks in the halo mass range $10^{11.5}-10^{12}$~$h^{-1}$~M$_{\odot}$ at all redshifts, with the peak evolving modestly from lower to higher halo masses from $z=5$ to $z=2$, and then being approximately constant for $z<2$.  The contribution from satellite galaxies spans a broader range of halo mass and peaks at higher halo mass, however, it is much smaller than that of the central galaxies, being only $\sim6\%$ of the total $350$~$\mu$m emissivity at $z=3.1$ and only $\sim14\%$ at $z=0.5$.

In Fig.~\ref{fig:SPIRE_Pk} we compare $P^{\nu}_{\rm I}({k_{\theta}})$ predicted by our model to the observations of \cite{Viero13}.  The horizontal dashed line in each panel represents the predicted shot noise.  This is the power that would be expected if the background were composed of an un-clustered population of point sources and as such has no scale-dependence.  It is related to the number counts of the model by
\begin{equation}
P^{\nu}_{\rm shot} = \int_{0}^{S_{\rm cut}}{S_{\nu}}^{2}\frac{{\rm d}\eta}{{\rm d}S_{\nu}}{\rm d}S_{\nu}{\rm ,}
\label{eq:shot_noise}
\end{equation}
\citep[e.g.][]{TegmarkEfstathiou96}, where $S_{\rm cut}$ is the limit above which sources can be resolved and are therefore removed/masked from further analysis in order to reduce the shot noise\footnote{Imposing the limit $S_{\rm cut}$ is necessary as for Euclidean number counts (${\rm d}\eta/{\rm d}S\propto S^{-2.5}$) the integral in equation (\ref{eq:shot_noise}) does not converge.}.  Note that this contribution to the power spectrum corresponds to the Dirac delta function term in equation (\ref{eq:wI_int}).  

We show the two extremes of masking schemes applied by Viero et al. to their data, in order to reduce the shot noise in their images.  They identified sources by finding peaks $>3\sigma$ in the matched-filtered SPIRE images at each wavelength.  Sources above a given flux limit ($S_{\rm cut}$) were then masked by circles with a $1.1\times\rm{FWHM}$ diameter, before calculating the power spectra.  Extended sources were removed by using the criterion $S_{\rm cut}=400$~mJy.  We compare to the most extreme masking case $S_{\rm cut}=50$~mJy (open squares) and mimic the masking applied by \cite{Viero13} by excluding galaxies with $S_{\nu}>50$~mJy prior to the creation of our simulated images.  We have tested that masking pixels in the full image produces near identical results.

At $350$ and $500$~$\mu$m we also compare our predictions to the observational data of the Planck Collaboration (XXX, \citeyear{Planck14CIB}). These authors employ a slightly different masking scheme to that used by Viero et al., however this has a negligible effect on the scales covered by their data.  Encouragingly, both observational datasets are in good agreement.    

We note that there is a discrepancy between the model predictions and the observational data of a factor $\sim 2$ over all wavelengths and angular scales. Whilst this represents much better agreement than for previous versions of our model \citep[e.g.][]{Hank12} we investigate whether it is possible to further improve this by forcing a better agreement between our predicted number counts and those that are observed.  By construction, this gives us the observed surface density of objects and should make the shot noise terms equal.  This is merely an illustrative exercise to replicate one of the freedoms of empirical models which are constrained to match the observed counts e.g. HOD modelling.  An example of this is shown in Fig.~\ref{fig:FluxScale}, where we scale the fluxes of our galaxies by the function shown in the bottom panel, chosen such that it brings our model number counts into better agreement with the observed data (top panel).  We then apply this scaling relation to our galaxies prior to the creation of our simulated images and recalculate the power spectrum, resulting in the dashed red line in Fig.~\ref{fig:SPIRE_Pk}.  This exercise produces power spectra in much better agreement with the observed data, even at low values of $k_{\theta}$ where clustering dominates over the shot noise.  We recognise that this is an artificial adjustment to our model.  However it is a relatively minor one as we do not adjust the flux of our galaxies by more than $\sim 40\%$ across all three bands.  We do not draw strong conclusions from this, but simply note that good agreement with the observed number counts is required to reproduce the observed power spectra.  In this case we have adjusted our number counts artificially but in future this could be achieved by developments to the treatment of physical processes in the model.     

At $250$~$\mu$m there remains a small ($\sim25\%$) discrepancy between the observed shot noise and that predicted by our flux rescaling, despite the fact that the number counts are in close agreement ($\sim14\%$).  We attribute this to field-to-field variation between the fields used to measure the observed number counts and those used for measuring power spectra, and the uncertainties on both measurements.

\begin{figure*}
\includegraphics[trim = 0 190 0 0,clip = True,width=0.9\linewidth]{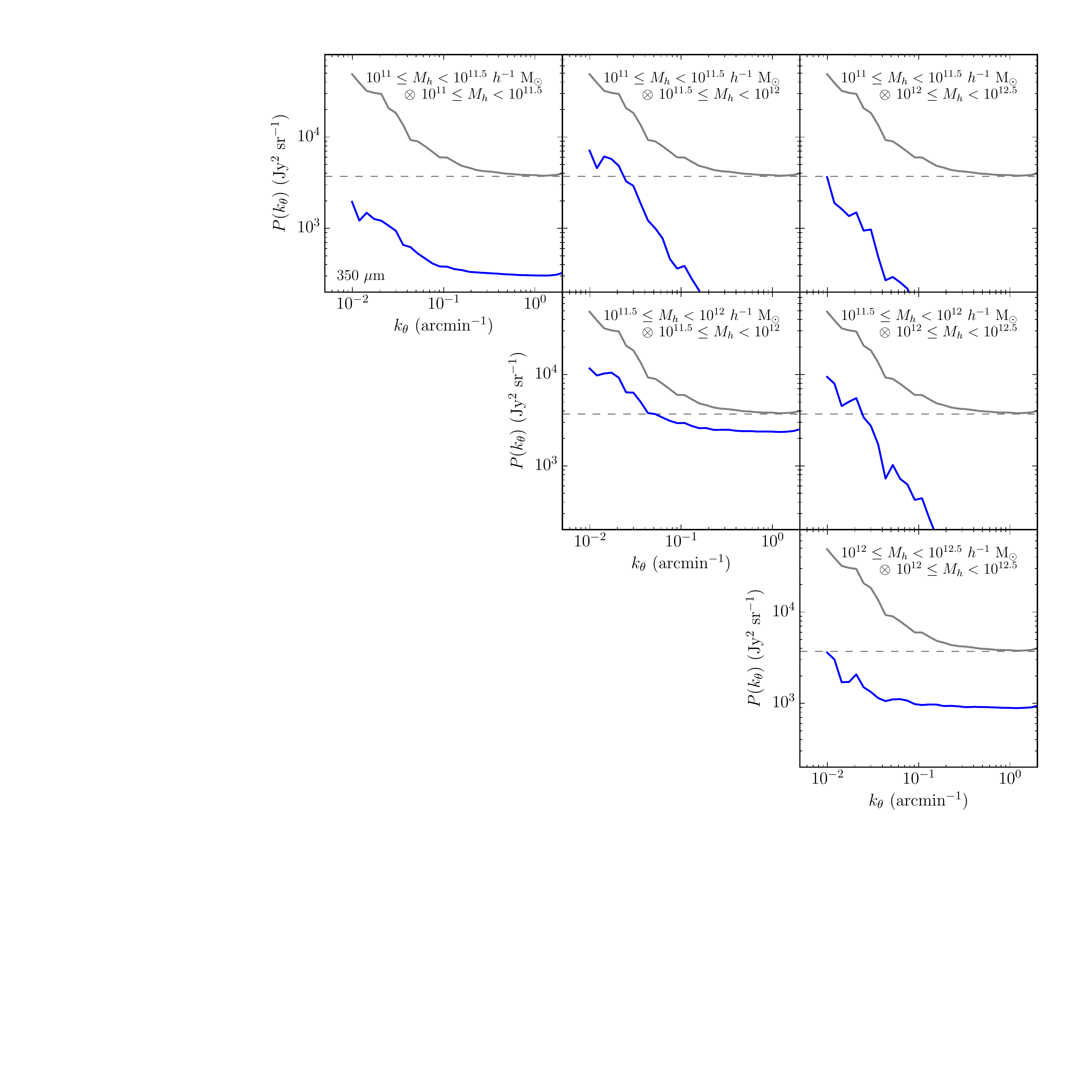}
\caption{Power spectrum of the CIB predicted by our model at $350$~$\mu$m for $S_{350\mu\rm m}<50$~mJy (solid grey line) divided into the following halo mass bins $10^{11} M_{\rm h}\leq10^{11.5}$~$h^{-1}$~M$_{\odot}$, $10^{11.5}\leq M_{\rm h}<10^{12}$~$h^{-1}$~M$_{\odot}$ and $10^{12}\leq M_{\rm h}<10^{12.5}$~$h^{-1}$~M$_{\odot}$.  The diagonal panels indicate the auto-power spectrum of the halo mass bin indicated in the panel.  The off-diagonal panels indicate the cross-power spectrum between different bins, as indicated in the panel.  The dashed grey horizontal line indicating the total shot noise for the $S_{350\mu\rm m}<50$~mJy population.}
\label{fig:SPIRE_Pk_mhalobins}
\end{figure*}

As the FIR emissivity is dominated by a halo mass range of $10^{11.5}-10^{12}$~$h^{-1}$~M$_{\odot}$ (e.g. at $350$~$\mu$m and $z=3.1$, $54$\% of the total emissivity comes from halos in this mass range) we investigate whether this mass range also contributes most to the angular power spectrum of CIB anisotropies.  We retain the masking flux limit of $S_{\rm cut}=50$~mJy from Viero et al. and divide our lightcone catalogue into three halo mass bins of $0.5$ dex width, which span the peak of the differential emissivity distribution shown in Fig.~\ref{fig:jnu}.  We then construct an image for each bin.  The cross-power spectra for these images are shown in Fig.~\ref{fig:SPIRE_Pk_mhalobins}.  We have ignored the contribution from halos outside the mass bins chosen for this plot, however, the bins chosen contribute $\sim90$\% of the total power spectrum (for $S_{350\mu\rm m}<50$~mJy).  We can see that the same halo mass bin which dominates the emissivity dominates the contribution to the power spectrum, as one might expect if the fractional cross-power spectrum term, $\mathcal{P}^{\nu}_{j}(k,M_{\rm h},M_{\rm h}^{\prime},z)$, in equation (\ref{eq:define_Pk_Mh}) is a smoothly varying function of halo mass, given the peaked nature of the $\mathrm{d}j_{\nu}/\mathrm{d}\log_{10}M_{\rm h}$ term.   
 
\begin{figure*}
\includegraphics[width=0.9\linewidth]{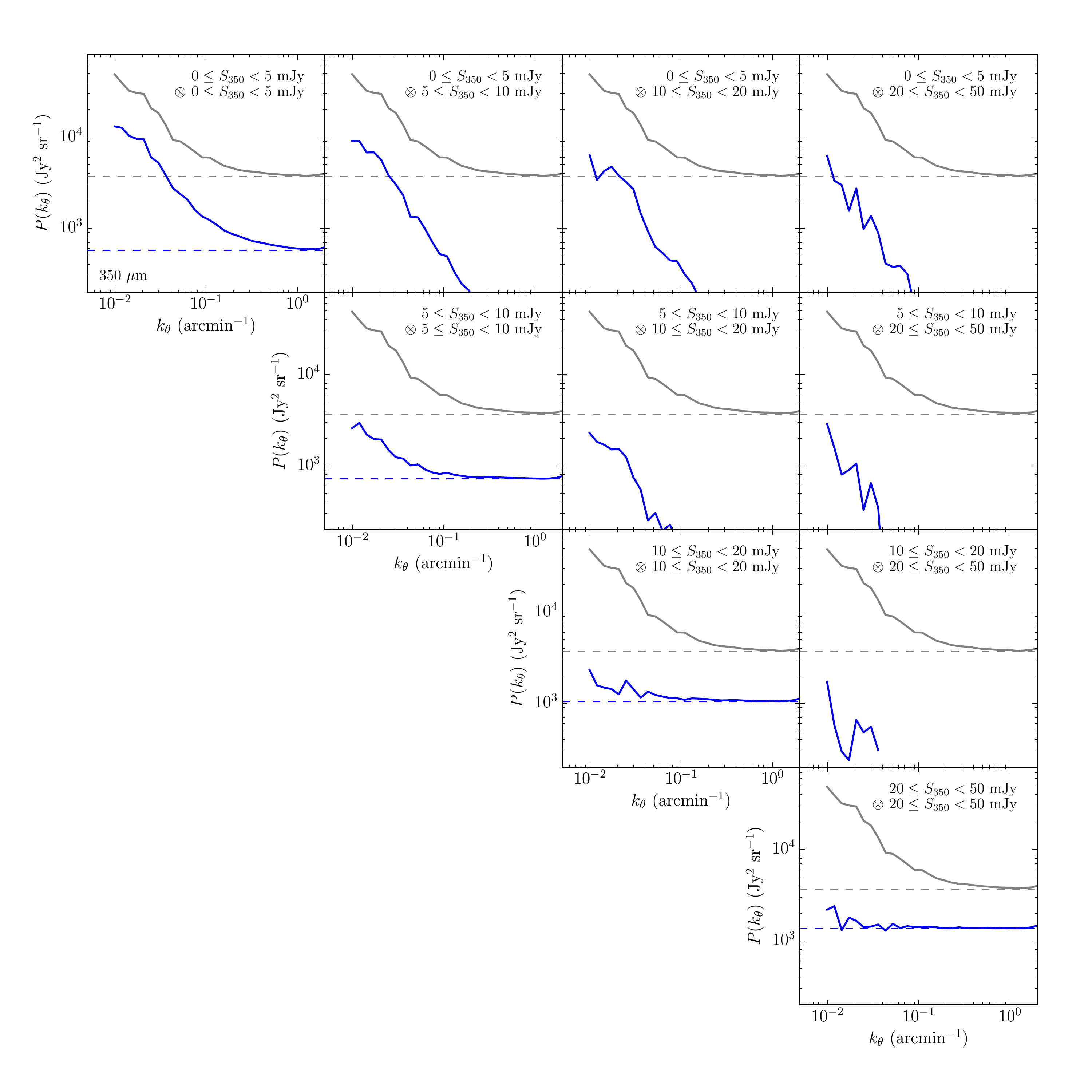}
\caption{Power spectrum of the CIB predicted by our model at $350$~$\mu$m divided into the following flux bins $0\leq S_{350\mu\rm m}<5$~mJy, $5\leq S_{350\mu\rm m}<10$~mJy, $10\leq S_{350\mu\rm m}<20$~mJy and $20\leq S_{350\mu\rm m}<50$~mJy.  The diagonal panels indicate the auto-power spectrum of the flux bin indicated in the panel and as such contains the shot noise term, indicated by the horizontal dashed line.  The off-diagonal panels indicate the cross-power spectrum between different bins, as indicated in the panel.  The solid grey line in each panel indicates the total power for $S_{\nu}<50$ mJy, with the dashed grey horizontal line indicating the total shot noise.}
\label{fig:SPIRE_Pk_Sbins}
\end{figure*}

To investigate the fluxes of the galaxies which contribute most to the power spectrum, we divide our lightcone catalogue into four flux bins and construct an image for each.  The cross-power spectra for these images shown for $350$~$\mu$m in Fig.~\ref{fig:SPIRE_Pk_Sbins}.  We can see immediately that on larger angular scales ($k_{\theta}\lesssim0.1$~arcmin$^{-1}$) the power is dominated by galaxies in the faintest bin $S_{\nu}<5$~mJy (e.g. top left panel), whilst the shot noise is dominated by brighter galaxies (e.g. bottom right panel).  In our model the dominant shot noise contribution at $350$~$\mu$m (for galaxies with $S_{350~\mu\rm m}<50$ mJy) comes from galaxies with $S_{350\mu\rm m}\sim20$~mJy.  
  
\section{Summary}
\label{sec:summary}
We present predictions for the clustering evolution of dusty star-forming galaxies selected by their total infra-red luminosity ($L_{\rm IR}$), and their emission at far infra-red (FIR) and sub-millimetre (sub-mm) wavelengths.  This includes the first predictions for potential biases on measurements of the angular clustering of these galaxies due to the coarse angular resolution of the single-dish telescopes used for imaging surveys at these wavelengths.  Our model incorporates a state-of-the-art semi-analytic model of hierarchical galaxy formation, a dark matter only $N$-body simulation which utilises the \emph{WMAP7} cosmology and a simple model for calculating the emission from interstellar dust heated by stellar radiation, in which dust temperature is calculated self-consistently.   

We present predictions for the spatial clustering of galaxies selected by the total infra-red luminosity for $L_{\rm IR}\sim10^{9}-10^{12}$~$h^{-2}$~L$_{\odot}$ for $z=0-5$.  We find that the clustering evolution in our model depends on the luminosity of the selected galaxies.  The large-scale bias evolution of our most luminous galaxies ($10^{12}-10^{12.5}$~$h^{-2}$~L$_{\odot}$) is consistent with them residing in halos of mass $10^{11}-10^{12}$~$h^{-1}$~M$_{\odot}$ over this redshift range.  In the model, this halo mass range is the one most conducive to star formation over these redshifts.  For lower luminosity populations the range of halo masses selected changes with redshift, such that generally they move to higher mass halos with increasing redshift.    

We find that $850$~$\mu$m selected galaxies in our model represent a clustered population, with an $S_{850\mu\rm m}>4$~mJy selected sample having a correlation length of $r_{0}=5.5_{-0.5}^{+0.3}$ $h^{-1}$~Mpc at $z=2.6$, consistent with observations of \cite{Hickox12} and \cite{Blain04}.  The bias with which they trace the dark matter evolves with redshift in a way consistent with the SMGs residing in halos of $10^{11.5}-10^{12}$~$h^{-1}$~M$_{\odot}$ up to a redshift of $z\sim4$.  This result is insensitive to the flux limit used to select the galaxies for $0.25\lesssim S_{850\mu\rm m}\lesssim4$~mJy, and we note that even at the faintest fluxes investigated ($S_{850\mu\rm m}\gtrsim0.25$~mJy) the model predicted $850$~$\mu$m number counts are dominated by starburst galaxies.  Interestingly, the halo occupation distribution for $850$~$\mu$m central galaxies peaks well below unity.  Halo abundance matching models which force the HOD of central galaxies to equal unity would place galaxies in much more massive halos than our model, given the same galaxy number density.  We find further that our brightest SMGs ($S_{850\mu\rm m}>4.0$~mJy) evolve into $z=0$ galaxies with stellar mass $\sim10^{11}$~$h^{-1}$~M$_{\odot}$, occupying a broad range of present day halo masses $10^{12}-10^{14}$~$h^{-1}$~M$_{\odot}$.  Thus, in our model, bright SMGs do not necessarily trace the progenitors of the most massive $z=0$ environments.  Our $S_{850\mu\rm m}$ selected galaxy populations share significant overlap with the most infra-red luminous galaxy populations $L_{\mathrm{IR}}\sim10^{12}$~$h^{-2}$~L$_{\odot}$, and thus exhibit similar clustering evolution.

We make predictions for the angular clustering of sub-mm sources identified in the SCUBA-2 Cosmology Legacy Survey.  We show that the angular clustering of $850$~$\mu$m single-dish selected sources is biased with respect to that of the underlying galaxy population, in our model by a factor of $\sim4$.  We attribute this `blending bias' to the coarse angular resolution of single dish telescopes blending the sub-mm emission of many (typically physically unassociated) galaxies into a single source.  This induces cross-correlation terms between sources selected at different redshifts. The position of a galaxy at $z_{\rm A}$ boosted into the source selection by fainter galaxies at some other redshift $z_{\rm B}$ will thus be correlated with the positions of galaxies at $z_{\rm B}$, some of which will already be included in the source selection.  It is the addition of these induced cross-correlations that leads to the `blending bias'.  The value of this bias depends on the size of the beam, the intrinsic clustering of the underlying galaxy population, and their number counts.  

We caution that this severely complicates the interpretation of measurements of the angular clustering of SMGs derived from single-dish survey source catalogues, and if not considered could lead to the halo masses for SMGs being significantly overestimated.  The angular clustering of galaxies selected at $850$~$\mu$m in our model is insensitive to the flux limit used (as is the case for the spatial clustering), and agrees with the angular clustering of intensity fluctuations predicted by the model at that wavelength.

The FIR emissivity of our model is dominated by the emission from halos in the mass range $10^{11.5}-10^{12}$~$h^{-1}$~M$_{\odot}$ independent of redshift, and this halo mass range also dominates the angular power spectrum of CIB anisotropies.  Our model agrees with the observed angular power spectrum of CIB anisotropies at \emph{Herschel}-SPIRE wavelengths ($250$, $350$ and $500$~$\mu$m, Viero et al. \citeyear{Viero13}) to within a factor of $\sim 2$ over all scales, representing an improvement over previous versions of the model. This agreement can be further improved on by making minor ($\lesssim40\%$) artificial adjustments to the fluxes of our galaxies which bring the predicted number counts into better agreement with those observed.

Galaxies selected by their FIR/sub-mm emission represent a large proportion of the cosmic star formation over the history of the Universe.  As such, understanding the nature of these galaxies is critical to a full understanding of galaxy formation.  In our model, the galaxies that contribute to the bulk of the CIB are predominantly disc instability triggered starbursts which reside in a relatively narrow range of halo masses $10^{11.5}-10^{12}$~$h^{-1}$~M$_{\odot}$ for $z\lesssim5$. 

Abundance matching arguments which combine the observed stellar mass function with the theoretically predicted halo mass function at $z=0$ imply that this is also the mass range for present-day halos for which the conversion of baryons into stars has been most efficient \citep[e.g.][]{Guo10}. The stellar fraction in a halo depends on an integral over the past history of star formation in all of the progenitors of that halo. In our model, the fact that the conversion efficiency of baryons into stars peaks in present day halos of mass $\sim 10^{11.5}-10^{12}$~$h^{-1}$~M$_{\odot}$ is a simple consequence of most of the star formation occurring in such halos over a large range of redshifts ($z\lesssim5$), combined with the growth of halos by hierarchical structure formation. This in turn is a consequence of the physical prescriptions on which our model for galaxy formation is based, in particular for gas cooling in halos and feedback from supernovae and AGN.  Observationally, information regarding the host halo masses of selected galaxies can be derived from measurements of their clustering, however extracting significant results from observations at FIR/sub-mm wavelengths is a challenging exercise.  This work presents predictions which we hope will inform the interpretation of future observations.       

\section*{Acknowledgements}
The authors wish to thank Lingyu Wang and Chian-chou Chen for helpful discussions, and the anonymous referee for providing a detailed and constructive report.  This work was supported by the Science and Technology Facilities Council [ST/K501979/1, ST/L00075X/1].  CMB acknowledges the receipt of a Leverhulme Trust Research Fellowship.  This work used the DiRAC Data Centric system at Durham University, operated by the Institute for Computational Cosmology on behalf of the STFC DiRAC HPC Facility (www.dirac.ac.uk). This equipment was funded by BIS National E-infrastructure capital grant ST/K00042X/1, STFC capital grant ST/H008519/1, and STFC DiRAC Operations grant ST/K003267/1 and Durham University. DiRAC is part of the National E-Infrastructure. 

\bibliographystyle{mn2e}
\bibliography{ref.bib}
\label{lastpage}
\end{document}